\documentclass[11pt,a4paper]{article}

\usepackage{amsmath,amssymb}
\usepackage{epsfig,graphicx}
\usepackage{subfigure}
\usepackage{graphicx}
\usepackage{rotating}
\usepackage{cancel}
\usepackage{bm}
\usepackage{color}
\usepackage{comment}
\usepackage{braket}
\usepackage{cite}
\usepackage{psfrag}
\usepackage{slashed}
\usepackage{soul}
\usepackage{tikz}
\usepackage{hyperref}
\usepackage{bbold}
\usepackage{wasysym}
\usepackage[normalem]{ulem}

\renewcommand\[{\left[}

\newcommand{\exclude}[1]{}

\def\beq{\begin{equation}}
\def\eeq{\end{equation}}

\begin{document}
\date{}

\title{
\huge{\textbf{Limits on New Lorentz-violating Bosons}}}

\author{P. Carenza$^1$, J. Jaeckel$^{2}$, G. Lucente$^{2,3}$, T. K. Poddar$^{4}$, N. Sherrill$^{5,6}$ and M. Spannowsky$^{7}$
\\[2ex]
\small{\em $^1$ The Oskar Klein Centre, Department of Physics, }\\ \small{\em Stockholm University, Stockholm 106 91, Sweden} 
\\[0.5ex]
\small{\em $^2$Institut f\"ur theoretische Physik, Universit\"at Heidelberg,} \\
\small{\em Philosophenweg 16, 69120 Heidelberg, Germany}\\
[0.5ex]
\small{\em $^3$ SLAC National Accelerator Laboratory,} \\
\small{\em 2575 Sand Hill Rd, Menlo Park, CA 94025}
\\[0.5ex]
\small{\em $^4$ INFN Sezione di Napoli, Gruppo collegato di Salerno,} \\
\small{\em Via Giovanni Paolo II 132 I-84084 Fisciano (SA), Italy,}\\
[0.5ex]
\small{\em $^{5}$Department of Physics and Astronomy, University of Sussex}\\
\small{\em Brighton, BN1 9QH, United Kingdom}\\[0.5ex]
\small{\em $^{6}$Institut f\"ur Theoretische Physik, Leibniz Universit\"at Hannover}\\
\small{\em Appelstra{\ss}e 2, Hannover, 30167, Germany}\\[0.5ex]
\small{\em $^{7}$Institute for Particle Physics Phenomenology, Department of Physics,}\\
\small{\em  Durham University, Durham DH1 3LE, United Kingdom}
}

\maketitle

\begin{abstract}
We obtain novel constraints on new scalar fields interacting with Standard Model fermions through Lorentz-violating couplings, bridging searches for scalar particles and Lorentz-symmetry tests. These constraints arise 
from torsion-balance experiments, magnetometer searches, and an excessive energy loss in Red Giant stars.
Torsion-balance experiments impose stringent constraints, benefitting 
from large macroscopic sources including the Sun and Earth. Magnetometer-based searches, which detect pseudo-magnetic fields through spin precession, offer additional limiting power to low-mass scalar fields. Meanwhile, observations of Red Giant stars place strong limits on additional energy loss mechanisms, extending these constraints to higher scalar mass ranges and a wider range of Lorentz-violating couplings.
Combining data from laboratory experiments and astrophysical observations, this approach strengthens constraints on Lorentz-violating interactions and paves the way for future investigations into physics beyond the Standard Model.

\end{abstract}

\newpage

\section{Introduction}
Searching for new physics has long relied on two primary avenues: the examination of fundamental symmetries and the search for new particles. Each approach has independently contributed to significant breakthroughs, including the discovery of parity violation in the weak interaction~\cite{Wu:1957my} and the subsequent identification of the $W$ and $Z$ bosons~\cite{GargamelleNeutrino:1973jyy,UA1:1983crd,UA1:1983mne}. These successes highlight the value of considering more complex scenarios incorporating elements of both symmetry violations and particle searches. In this paper, we will follow this approach, considering Lorentz-symmetry violations in interactions with a new scalar particle.

Lorentz symmetry is a cornerstone of the Standard Model (SM) of particle physics and General Relativity. Precision tests have demonstrated that Lorentz symmetry and the associated CPT (Charge conjugation-Parity-Time reversal) symmetry are upheld with extraordinary accuracy~\cite{Kostelecky:2008ts}\footnote{For a phenomenological  investigation into possible violations of the larger Poincare symmetry group see~\cite{Gupta:2022qoq}.}. However, these tests have not covered all energy and length scales, nor have they exhausted the multitude of routes these symmetries could be violated. Any deviation observed in suitable precision experiments from the standard predictions would point to new physics, challenging the universality of these foundational symmetries.

The Standard-Model Extension (SME) \cite{Colladay:1996iz, Colladay:1998fq,Kostelecky:2003fs} framework provides a systematic parametrization for incorporating violations of Lorentz symmetry. This framework constructs all possible tensorial operators involving quantum and gravitational fields using an effective field theory approach. The coefficients of these new operators quantify the type and degree of Lorentz Violation (LV). Extensive studies have explored the stability \cite{Kostelecky:2000mm,Kostelecky:2024rsn} and renormalizability \cite{Kostelecky:2001jc,Colladay:2007aj,Colladay:2009rb} of this extension, ensuring its theoretical consistency. Potential theories for Lorentz-symmetry violation include spontaneous symmetry breaking in string theory \cite{Kostelecky:1988zi,Kostelecky:1991ak,Altschul:2005mu}, phenomena in loop quantum gravity \cite{Gambini:1998it,Alfaro:2001rb}, variations in fundamental constants over spacetime \cite{Kostelecky:2002ca,Ferrero:2009jb}, and non-commutative geometry \cite{Mocioiu:2000ip,Carroll:2001ws}, among others. 

As already mentioned, the effects of LV have been probed with amazing precision~\cite{Kostelecky:2008ts}. However, most studies have been performed in the context of the SME, building on an effective field theory approach with {\emph{no new particles}} within the accessible energy range. This limits interaction effects to be point-like and precludes effects from the production of new particles. Here, we want to explore both of these aspects, considering long-range interactions as well as the production of the new particles in stars leading to an additional energy-loss channel.

Concretely, an intriguing yet less extensively studied aspect 
is the possibility that the LV occurs (only) in interactions of a new, light scalar with SM fermions. 
At low energies, such interactions give rise to the Yukawa potentials and their derivatives, which are altered in the presence of Lorentz-violating effects. The theoretical foundation for such models has been outlined in~\cite{Altschul:2006jj,Ferrero:2011yu,Altschul:2012xu}.  Our study employs this framework to derive explicit constraints on a new scalar field coupled to SM fermions through Lorentz-violating interactions, with a particular focus on mass-dimension-4 terms in the Lagrangian.
We consider potentials~\cite{Altschul:2012xu} arising from the virtual exchange of the scalar boson between test masses in an Earth-bound experiment and sources within the solar system and beyond. In addition, we are also considering scalar emission in scattering effects in an astrophysical environment, such as in Red Giant (RG) stars.

This approach allows for a systematic exploration of potential new physics by interfacing with both particle phenomenology and tests of fundamental symmetries. Concretely, we start from a Lagrangian \cite{Altschul:2012xu}, 
\begin{align}
\label{YLV}
&{\cal L}_{\rm Y} = \frac{1}{2}(\partial_{\mu}\phi)^2-\frac{1}{2}m^2_\phi\phi^2- \phi\sum_{F} \bar\psi_{F} G_{F}\psi_{F},  \\\nonumber
& G_{F} = g_{F} + ig'_{F}\gamma_5 + I^{F}_\mu \gamma^\mu 
+ J^{F}_\mu \gamma_5\gamma^\mu + \frac{1}{2}L^{F}_{\mu\nu}\sigma^{\mu\nu},
\end{align}
 where $m_\phi$ denotes the mass of the scalar field $\phi$, $g_F$ and $g'_F$ are the conventional scalar and pseudoscalar couplings, respectively. The sum $F$ runs over the SM fermion species. The dimensionless coefficients, parametrizing vector, pseudovector and second-rank tensorial coefficients for LV are $I^{F}_\mu, J^{F}_\mu$, and $L^{F}_{\mu\nu}$. By assumption, these terms describe perturbative Lorentz-violating effects around $g_F$ and $g^\prime_F$. Note that the vector ($I^{F}_\mu$) and pseudovector ($J^{F}_\mu$) coefficients also induce violations of CPT symmetry because the associated field operators are odd under CPT conjugation. 
The Yukawa interaction terms in Eq.~\eqref{YLV} significantly impact the interaction potential between fermions, yielding unique signatures in non-relativistic and relativistic limits. This work explores these scenarios by examining how low-mass scalars ($m_\phi \lesssim  {\mathcal{O}}(10)$\;keV) interacting with SM fermions can be constrained by laboratory experiments and astrophysical observations. To our knowledge, so far only a single study has assessed the sensitivity to the coefficient combination $I_0^F+J_0^F$ from pions decaying to muons and neutrinos, yielding constraints at the $10^{-10}$ level ~\cite{Altschul:2014gqa}. 

\begin{figure}
\centering
\includegraphics[width=\linewidth,trim=53 10 53 10,clip]{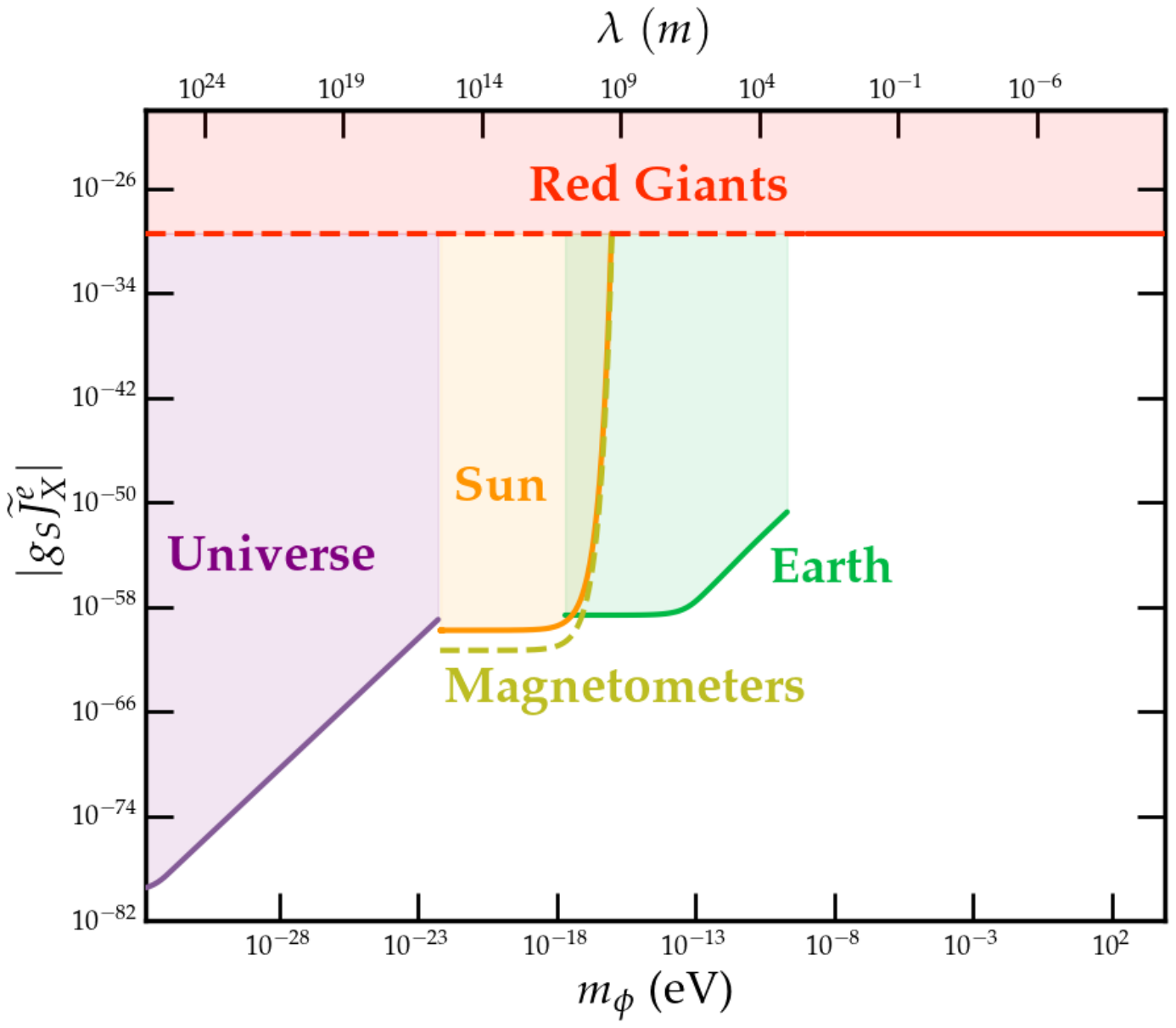}
\caption{
Constraints on $|g_S\widetilde{J}^e_X|$ (cf.~Eqs.~\eqref{YLV} and \eqref{tildebasis}) 
as a function of the mediator mass $m_\phi$ (bottom) and wavelength $\lambda$ (top).
Constraints on the $Y$ and $Z$ components are qualitatively similar on this double logarithmic plot over many orders of magnitude, but can quantitatively change by up to a factor of $\sim 100$ (see Table~\ref{tab1}). 
The depicted limits for the potential tests and the sensitivity for the magnetometers are for the source coupling, $g_S$, being a coupling to nucleons. However, the limits for a source coupling to electrons are quite similar. The same holds for a Lorentz-violating coupling of the test mass to nucleons, except in the magnetometer case. 
The astrophysical (RG) constraints shown in red are for $g_S=g_e$ only. The RG bound is shown as a dashed line for $m_a\lesssim10^{-9}$~eV ($\lambda\gtrsim 100$~m) since for lower masses the long-range force induced by the boson may compete with gravity. For more details see the text in the following sections.
\label{fig:results}}
\end{figure}

Our investigation begins in Sec.~\ref{sec:potentials} by looking at long-range forces from the exchange of the light scalar. In particular we analyze torsion-balance experiments \cite{Heckel:2006ww,Heckel:2008hw,Bluhm:1999ev}, which have been instrumental in testing potential fifth forces and deviations from known physics \cite{Hees:2018fpg,Adelberger:2003zx,Fischbach:1996eq,KONOPLIV2011401,Berge:2017ovy,Brzeminski:2022sde}.
We also explore the sensitivity of magnetometer-based searches  \cite{PhysRevLett.89.253002,Kornack:2005xrn,Brown:2010dt,Hoedl:2011zz,Peck:2012pt,Allmendinger:2013eya,Budker:2013hfa,Kahn:2016aff,Garcon:2017ixh,Crescini:2017uxs,Lee:2018vaq,Chu:2018dgp,Kim:2021eye,Bloch:2022kjm,Agrawal:2023lmw}. These highly sensitive devices measure spin-precession frequencies, providing a powerful tool for detecting spin-dependent interactions with external fields.
In Sec.~\ref{sec:redgiant}, we extend our analysis to astrophysical settings, mainly focusing on energy-loss processes in RG stars. Due to their large volume, high density and temperature, such stellar environments are known to be sensitive to energy loss caused by the production of new low-mass particles~\cite{Raffelt:1994ry,Capozzi:2020cbu,Straniero:2020iyi,Carenza:2021osu}. We use these systems to probe Lorentz-violating new scalars with masses up into the keV-range.
Our results are summarized in Figure.~\ref{fig:results}. We offer concluding remarks in Sec.~\ref{sec:conclusions}.

\section{Probing scalar-mediated long-range Lorentz-violating potentials}\label{sec:potentials}

For low-mass bosons, searches for new forces are usually a powerful tool (cf, e.g.~\cite{Adelberger:2009zz} for a review and ~\cite{Moody:1984ba,Dobrescu:2006au} for some examples of early theory papers).  
In the non-relativistic limit, the fermion–fermion scattering process described by Eq.\eqref{YLV} leads to Lorentz-violating corrections to the Yukawa potential and its derivatives\cite{Altschul:2012xu}.

We follow  Ref.~\cite{Altschul:2012xu} by assuming LV only occurs at the vertex of fermion $A$ with mass $m_A$, while the second fermion $B$ with mass $m_B$ is conventional in the sense of having only scalar and pseudoscalar couplings $g_B$ and  $g'_{B}$, respectively. 

Let us briefly comment on why this simplification is justified in the cases of interest to us. To probe spin-dependent Lorentz-violating long-range potentials, torsion-balance and magnetometer experiments are ideal setups, with the experimental test mass consisting of $A$-type fermion spins.
The source of the potential is taken as being composed of $B$-type fermions, which have a coupling with $A$-type fermion spins. For a macroscopic object (such as a star or a planet) consisting of $B$-type fermions, the net polarization is zero, and therefore, most of the Lorentz-violating effects (which are spin-dependent) cancel out. 
The only exception is $I$-dependent terms whose vectorial components are also velocity suppressed, which, in the following, we anyway will not constrain.   
Furthermore, we implicitly assume a hierarchy that the Lorentz-violating effects are small compared to the ordinary interactions. Second-order Lorentz-violating effects are hence further suppressed.
All in all, treating only $A$ as Lorentz-violating is a reasonable approximation.

Under these assumptions there are two contributions\footnote{See also Appendix \ref{app1} for the derivation details of the Lorentz-violating scalar-pseudovector potential.}~\cite{Altschul:2012xu},
\begin{align}
\label{pots}
&V_f(r) = \left[\widetilde{g}_{A} - I^{A}_j(v^{A}_{\rm av})_j - J^{A}_0(v_{\rm av}^{A})_j\sigma_j^{A} + \widetilde{J}^{A}_j\sigma_j^{A} + 
\epsilon_{jkl}\widetilde{L}^{A}_j(v_{\rm av}^{A})_k\sigma_l^{A} \right]g_{B} f(r), \\ \nonumber
&V_g(\vec{r}) = \frac{g_{B}}{2m_{A}}\left[g'_{A}\sigma_j^{A} - \widetilde{L}^{A}_j -\epsilon_{jkl}I^{A}_k\sigma_l^{A}\right]g_j(\vec{r})  \\
&\hspace{.9cm} -\frac{g'_{B}}{2m_{B}}\left[\widetilde{g}_{A} - I^A_k(v_{\rm av}^{A})_k-J^{A}_0(v_{\rm av}^{A})_k\sigma^{A}_k + \widetilde{J}^{A}_k\sigma^{A}_k + \epsilon_{kln}\widetilde{L}^{A}_k(v_{\rm av}^{A})_l\sigma_n^{A}\right]\sigma_j^{B} g_j(\vec{r}),
\label{potss}
\end{align}
to the the total non-relativistic potential $V_f(r) + V_g(\vec{r})$. Here, $\vec{v}^{A}_{\rm av}$ is the average velocity of the scattered fermions at the Lorentz-violating vertex. Additionally, in Eqs.~\eqref{pots}-\eqref{potss} we have introduced
\begin{align}
\label{fg}
    &f(r) = -\frac{e^{-m_\phi r}}{4\pi r},\\
    &g_j(\vec{r})=\partial_j f(r)= \frac{e^{-m_\phi r}}{4\pi}
    \left(\frac{m_\phi}{r} + \frac{1}{r^2}\right)\frac{x_j}{r}
    \equiv g(r)\frac{x_j}{r} \,,
\end{align}
and expressed the potentials $V_{f}(r)$ and $V_{g}(\vec{r})$ in the coefficient basis
\begin{align}
\label{tildebasis}
&\widetilde{g}_{F} = g_{F} + I^{F}_0, \\\nonumber
&\widetilde{J}^{F}_j = J^{F}_j + \epsilon_{jkl}L^{F}_{kl}, \\ \nonumber
&\widetilde{L}^{F}_j = L^{F}_{0j} = -L^{F}_{j0}.
\end{align}

\bigskip 

In the following, we exploit torsion-balance and magnetometer experiments to probe the Lorentz-violating potentials introduced above. Indeed, these experiments are sensitive to long-range interactions and ideally suited for probing the implications of Lorentz-violating potentials, particularly those involving large-scale sources like the Sun and Earth. We discuss how such sources can significantly affect the constraints on scalar boson-mediated Lorentz-violating interactions due to their macroscopic nature and proximity to experimental setups. Laboratory setups are particularly valuable for studying long-range spin-dependent interactions because they allow the controlled creation of spin-polarized systems, enabling observing net spin effects on the coupling due to distant sources.

\subsection{Constraints on $g\widetilde{J}_j$ from torsion balances}\label{sec:torsion}

Let us now consider experimental results from torsion balances constraining specific coupling coefficients.
The E{\"o}t-Wash group~\cite{Heckel:2008hw} has performed a time-dependent
analysis of the spatially independent potential 
\begin{equation}
\label{Ve}
V_e = -\boldsymbol{\sigma}^e\cdot \boldsymbol{\widetilde{b}}^e,
\end{equation}
where $\widetilde{b}_j^e$ are a combination
of the spatial components of the coefficients for LV in the electron sector coupling to the electron's spin $\sigma_j^e$~\cite{Bluhm:1999ev}. 
Constraints on the three spatial components 
$\widetilde{b}_X^e, \widetilde{b}_Y^e$, and $\widetilde{b}_Z^e$
were placed at the $10^{-31}$\;GeV level, where the subscripts $J =  X, Y$, and $Z$ denote Cartesian equatorial or, equivalently in our case of interest, Sun-Centered Frame (SCF) coordinates~\cite{Bluhm:2001rw,Kostelecky:2002hh,Bluhm:2003un}.  The SCF is defined by a $Z$ axis parallel to the Earth's rotation axis and $X, Y$ axes defining the equatorial plane. The temporal coordinate $T$ is reference to the year 2000 vernal equinox. 
The constraints are reproduced in Table~\ref{tab1}.
These results permit, under a proper interpretation, 
the extraction of 
constraints on a subset of coefficients 
from Eq.~\eqref{pots}. 

In Ref.~\cite{Heckel:2008hw}, 
E{\"o}t-Wash also examined the signatures of 
several Lorentz-invariant potentials with
spatial dependence controlled by $f(r)$ and
$g(r)$ defined in Eq.~\eqref{fg}. Their 
torsion-balance apparatus with polarized electrons ($e$)
was treated as a test mass in several source potentials
generated by a large number of nucleons ($N$). 

Comparing with Eq.~\eqref{pots}, we can see that in our case the  Lorentz-violating
potential has $A=e$ and $B=N$ fermion indices
and is controlled by the 
$\widetilde{J}$-type coefficients
defined in Eq.~\eqref{tildebasis}, 
\begin{equation}
\label{VJpointgeneral}
V_{\widetilde{J}}(\vec{r}) = 
\left(\boldsymbol{\sigma}^e\cdot \boldsymbol{\widetilde{J}}^e\right)
\left[g_N f(r) - \frac{g_N'}{2m_N}
(\boldsymbol{\sigma}^N\cdot \boldsymbol{\widehat{r}})g(\vec{r})\right].
\end{equation}
The electron spins have a Lorentz-violating coupling and the resulting interaction potential is sourced by nucleons. In Eq.~\eqref{VJpointgeneral},
the unit vector $\boldsymbol{\widehat{r}}$ 
points from the E{\"o}t-Wash apparatus of polarized electrons
to an infinitesimal nucleon 
source element of density $\rho(\vec{r})$. 
Consistent with what we argued above, existing results are not sensitive to the polarization
of the source nucleons $\boldsymbol{\sigma}^N$ as none of the considered
sources contain a significant number of polarized nucleons\footnote{Macroscopic sources could have a small amount of polarized electrons due to the spherical symmetry breaking of the sources or statistical fluctuation due to a magnetic field of the sources \cite{Hunter:2013hza,Clayburn:2024zxx,Poddar:2023bgk}.}. Accordingly, 
the second piece within the brackets of 
Eq.~\eqref{VJpointgeneral} does not contribute. Therefore,
\begin{equation}
\label{VJpoint}
V_{\widetilde{J}}(\vec{r}) \rightarrow
\left(\boldsymbol{\sigma}^e\cdot\boldsymbol{\widetilde{J}}^{e}\right)
g_N f(r)\,.
\end{equation}
This is simply an $f(r)$-dependent form of $V_e$ in Eq.~\eqref{Ve} and can be directly compared to the terms in Eq.~\eqref{pots}. 

So far the function $f(r)$, given in Eq.~\eqref{fg}, describes a 
Yukawa potential function  between two point-like
fermions separated by $r$. This must be generalized
to the potential between a macroscopic object, the
source of unpolarized nucleons, 
and an approximately point-like object, the apparatus.
This can be done by using the source integral
\begin{equation}
\label{eq:originalsourceint}
f(r) \rightarrow
-\frac{1}{4\pi u}
\int d^3r' \frac{\rho(\vec{r}')
}{|\vec{r}-\vec{r}'|}
e^{-|\vec{r}-\vec{r}'|/\lambda},
\end{equation}
where $u$ is the atomic mass unit
and $\lambda = 1/m_\phi$ is the wavelength
of the boson mediator.
Integrating over a source
introduces a mass scale.
We normalize $f(r)$ to $u\approx m_N$ to facilitate comparison with the point-source approximation and retain the same units.

Consider a source of radius $r_0$ and mass $M$ at a distance $R_0$. Assuming spherical symmetry and constant density $\rho(\vec{r}') =\rho_0$, where
\begin{equation}
\rho_0\approx \frac{3 M}{4\pi r_0^3}\, ,
\end{equation}
the source integral in Eq.~\eqref{eq:originalsourceint} evaluated 
at a separation $r=R_0$ gives 
\begin{equation}
\label{sourceint}
-\frac{M}{4\pi u R_0}\left[\frac{\xi \cosh\xi - \sinh\xi}{\xi^3/3}\right]
\exp\left[-\frac{R_0}{\lambda}\right].
\end{equation}
The parameter $\xi = r_0/\lambda$ accounts for the point-source approximation. Indeed, when the wavelength of the mediator significantly exceeds the source size ($\xi \ll 1$), the source effectively behaves as a point particle and Eq.~\eqref{sourceint} reduces to
\begin{equation}
\label{eq:ptsrcint}
-\frac{M}{4\pi u R_0}
\exp\left[-\frac{R_0}{\lambda}\right].
\end{equation}
The same result for the potential function would be obtained by plugging in Eq.~\eqref{eq:originalsourceint} the point-source mass density $\rho(\vec{r})=M\delta^3(\vec{r})$. We mention here that a similar discussion is valid also for the constraints from magnetometer searches in Sec.~\ref{sec:magnetic}.

In order to illustrate general features of the Lorentz-violating couplings, we perform a phenomenological analysis largely following the assumptions of Refs.~\cite{Heckel:2006ww,Heckel:2008hw} by explicitly considering the Sun and the Earth as sources\footnote{In principle one could also use the Moon. However, the resulting potential is completely overshadowed by the Sun and the Earth potential. Therefore, we do not consider the Moon.}. We assume a uniform-density approximation for the Sun, where Eq.~\eqref{sourceint} applies, while we account for variations in Earth's density by using a standard density parametrization~\cite{Dziewonski:1981xy}, given the proximity of the experimental apparatus to the source.

Let us first consider the source to be the Sun, with radius $r_0 =  r_\odot\approx 7\times10^5\,{\rm km}$, mass $M=M_\odot\approx 2\times 10^{30}\,{\rm kg}$~\cite{astronomicalalmanac2024,sunwiki} and uniform density $\rho_0 = \rho_{\odot}$. The approximate separation between the 
apparatus and the Sun is 
$R_0 = R_\odot \approx$ 1 AU\footnote{This is the 
approximate distance between 
the centre of the Earth and the centre
of the Sun. This distance varies $\sim 2\%$ over the year. 
As assumed by Ref.~\cite{Heckel:2008hw}, we ignore these variations.}. In this context, the Sun can be well approximated as a point source, for which Eq.~\eqref{eq:ptsrcint} is valid. This expression reaches its maximum in the massless-boson limit, where $\lambda \rightarrow \infty$. Indeed, for $\lambda \gg r_\odot, R_\odot$, the exponential factor tends to unity and the strength of the Lorentz-violating coupling is primarily determined by the ratio $M/R_0$. In this limit, the comparison 
\begin{equation}
\label{mapping}
g_N\boldsymbol{\widetilde{J}}^{e}
\left(\frac{M_\odot}{4\pi u R_\odot}\right)
\leftrightarrow 
\boldsymbol{\widetilde{b}}^{e} 
\end{equation}
can be made with the coefficients $\boldsymbol{\widetilde{b}}^e$ from Eq.~\eqref{Ve}.

As mentioned, for the Earth ($M_{\oplus}\approx6\times 10^{24}\,{\rm kg}$~\cite{astronomicalalmanac2024,earthwiki}) the uniform-density approximation cannot be carried over due to significant density variations and its closeness to the experimental apparatus. Instead, we directly integrate Eq.~\eqref{eq:originalsourceint} along with the density parametrization of Ref.~\cite{Dziewonski:1981xy}. 

Note that, though one could consider the 
detailed composition of the source as done, e.g., in Ref.~\cite{Dzuba:2024pri}, we follow Ref.~\cite{Heckel:2008hw} in assuming equal couplings for nucleons ($g_N = g_p = g_n$) and $g_e = 0$ for the Sun. 
With the constraints on $\widetilde{b}_J^e$ in Table~\ref{tab1}, we find the limits for the Sun and Earth depicted as the orange and green regions in Figure~\ref{fig:results}, respectively.
New parameter space from mass scales $m_\phi \approx  10^{-10}~\mathrm{eV}$ through the
massless limit is probed\footnote{We cut off the constrained region at a distance of 1~km due to uncertainties about the details of the geometry and composition of the apparatus.  This should suffice for our order of magnitude accuracy goal. Since the effects we are considering are generally additive no cancellations are expected and one could put the cutoff even lower. However, if aiming for a higher precision than a factor of a few one should follow~\cite{Heckel:2008hw} and already cut off at 1000~km.}. The curves flatten at a magnitude controlled by $M/R_0$ and remain valid through the massless limit. 
The parameter space covered by ``Universe" (purple) takes over around $m_\phi \sim 10^{-22}$\;eV and is discussed in the following section.
Notably, incredibly small values of the coupling product ($\gtrsim 10^{-60}$) can be tested as shown in Table~\ref{tab1}, where with Eq.~\eqref{mapping}, one-sigma upper limits on $|g_N\widetilde{J}_J^{e}|$ are presented and valid for masses $m_\phi\lesssim 10^{-18}\,{\rm eV}$.

\renewcommand{\arraystretch}{1.25}
\begin{table}[t!]
\centering
\begin{tabular}{c c c c c}
\hline\hline
Parameter & Constraint 
(this work) & Parameter & Constraint (Ref.~\cite{Heckel:2008hw}) \\
\hline
$|g_N\widetilde{J}^e_X|$ & $< 2\times 10^{-60}$ &  $\widetilde{b}^e_X$ & $(-0.67\pm 1.31)\times 10^{-31}$\;GeV\\
$|g_N\widetilde{J}^e_Y|$ & $< 1\times 10^{-60}$ &  $\widetilde{b}^e_Y$ & $(-0.18\pm 1.32)\times 10^{-31}$\;GeV\\
$|g_N\widetilde{J}^e_Z|$ & $< 4\times 10^{-59}$ &  $\widetilde{b}^e_Z$ & $(-4\pm 44)\times 10^{-31}$\;GeV\\
\hline\hline
\end{tabular}
\caption{Constraints on $|g_N\widetilde{J}_J^{e}|$, assuming the Sun as a source. Constraints on $\widetilde{b}^e_J$ from Ref.~\cite{Heckel:2008hw} are given at 68\% C.L.}
\label{tab1}
\end{table}

\bigskip

Using the same sources, we can also obtain limits on the electron coupling of the source. Assuming the number of nucleons and electrons to be roughly equal up to a factor of order $1$ (based on charge neutrality), the limits are essentially the same.
Moreover, using the data from Ref.~\cite{Brown:2010dt} we can also set constraints on a similar level for Lorentz-violating test body interactions with nucleons, cf. Eq.~\eqref{tab1}.

\subsection{Constraints from magnetometer searches}\label{sec:magnetic}

The leading-order Lorentz-violating scalar potential term, Eq.~\eqref{VJpoint}, that can be searched for with the torsion balances can also be detected using magnetometers based on spin interactions (see Ref.~\cite{Kim:2021eye} for some example proposals).
This can be understood by comparing the Lorentz-violating interaction, Eq.~\eqref{VJpoint}, to
the Hamiltonian $\mathcal{H}$ of a particle $f$ with spin $\boldsymbol{\sigma}_f$ in a magnetic field $\mathbf{B}$~\cite{Kim:2021eye},
\begin{equation}\label{a0}
\mathcal{H}=-\mu_f\mathbf{\boldsymbol{\sigma}}_f\cdot\mathbf{B}.
\end{equation} 
Here, the particle's magnetic moment is given by $\mu_f\boldsymbol{\sigma}_f$.

Following this analogy, the unpolarized nucleons in the source, e.g. the Sun, generate a pseudo-magnetic field acting on electron spins,
 \begin{equation}B^{\mathrm{LV}}=\frac{g_S|\boldsymbol{\widetilde{J}}^{e}| M}{\mu_e 4\pi u R}\left[\frac{\xi \cosh\xi - \sinh\xi}{\xi^3/3}\right]
\exp\left[-\frac{R}{\lambda}\right]\sim \frac{g_S|\boldsymbol{\widetilde{J}}^{e}| M}{\mu_e 4\pi u R}
\exp\left[-\frac{R}{\lambda}\right]\,.
     \label{eq:m2}
 \end{equation}
Here, $g_S$ is the coupling of the source particles, $\mu_e$ represents the magnetic moment of the electron, $M$ is the mass of the source, $R$ denotes the distance between the source and the Earth-based apparatus, and $\xi=r/\lambda$ quantifies the size of the source relative to the range of the interaction. The simplification in the last step of Eq. \eqref{eq:m2} corresponds to the point source approximation or $\xi\ll 1$, where the wavelength of the boson is large compared to the size of the source.

We stress again that the magnetic field in 
Eq.~\eqref{eq:m2} does not correspond to a real magnetic field generated through electromagnetic interactions. Instead, this field arises from the Lorentz-violating interaction and only acts on the spins of the particles in this interaction, in our case electrons. Therefore, it can only be detected in magnetometers that are based on the spin interaction of these particles. Nevertheless, for comparison with existing magnetometers, the magnetic field analogy is useful.

To see how sensitive such a setup could be, let us consider the Sun as a source object.
The induced magnetic field, generated by the nucleons in the Sun, is given by,
\begin{equation}
    B^{\rm LV}\sim 2\,{\rm aT}\left(\frac{g_{S}|\boldsymbol{\widetilde{J}}^{e}|}{10^{-60}}\right) .
\end{equation}

In the presence of such a pseudo-magnetic field, the electron spins precess just like in an ordinary magnetic field, see~\cite{Kim:2021eye}. 
Importantly, the pseudo-magnetic field penetrates magnetic shielding and therefore 
can be measured in a magnetically shielded environment where ordinary magnetic fields are blocked~\cite{Kim:2021eye}.
In addition, the annual modulation of the signal can be 
used as an additional discriminator~\cite{Kim:2021eye},
detected using highly sensitive magnetometers. 
Finally, the signal can be measured more precisely by positioning magnetometers at various locations around the Earth, as done in experiments like GNOME \cite{Afach:2021pfd}. The sensitivity of $\mathcal{O}{\mathrm{(aT)}}$ with electron spin can be achieved with $^{3}\mathrm{He}-\mathrm{K}$ co-magnetometers \cite{Lee:2018vaq}. The QUAX-$g_pg_s$ experiment \cite{Crescini:2017uxs}, which probes scalar $(g_s)$--pseudoscalar $(g_p)$ interactions, utilizing a DC SQUID magnetometer, is also capable of detecting electron spin with comparable sensitivity.
Assuming a final magnetic-field sensitivity $B_{\rm sensitivity} \sim 2\,\mathrm{aT}$~\cite{Agrawal:2023lmw,Crescini:2017uxs,Kim:2021eye,Lee:2018vaq,Hoedl:2011zz}\footnote{A sensitivity of $\sim 200\,{\rm aT}$ has already been achieved in~\cite{Lee:2018vaq}.},
magnetometers would be able to probe values of the Lorentz-violating coupling
 \begin{equation}
     |g_S\widetilde{J}^e_{J}|\sim 1\times 10^{-60}\left(\frac{B_{\rm sensitivity}}{2\,{\rm aT}}\right),
      \label{eq:m3}
 \end{equation}
for masses $m_\phi\lesssim 1.3\times 10^{-18}~\mathrm{eV}$, determined by the Earth-Sun distance. Hence, the magnetometer-based experiments can serve as complementary probes for Lorentz-violating interactions. The sensitivity curve in probing Lorentz-violating coupling with magnetometers has been indicated by a gold dashed line in Figure~\ref{fig:results}.

\bigskip

In a similar way, one can also use experiments sensitive to nuclear spins (see~\cite{PhysRevLett.89.253002,Kornack:2005xrn,Brown:2010dt,Kim:2021eye,Allmendinger:2013eya,Peck:2012pt}). In this case we can use already performed experiments with K-$^{3}$He~\cite{Brown:2010dt}, $^{199}$Hg-$^{133}$Cs~\cite{Peck:2012pt} and $^{3}$He-$^{129}$Xe~\cite{Allmendinger:2013eya} co-magnetometers, to set actual bounds. The resulting limits on nuclear couplings, which are even somewhat stronger, can be found in Table~\ref{tab2}\footnote{Note that, for the same magnetic field sensitivity, the limits are stronger because of the much smaller nuclear magnetic moment to be used in Eq.~\eqref{eq:m2}.}.

\renewcommand{\arraystretch}{1.25}
\begin{table}[h]
\centering
\begin{tabular}{c c c c c c}
\hline\hline
Parameter & Constraint 
(this work) & Parameter & Constraint & Ref. \\
\hline
$|g_N \widetilde{J}^p_\perp|$ & $< 5\times 10^{-61}$ &  $\widetilde{b}^p_\perp$ & $<6 \times 10^{-32}$\;GeV & ~\cite{Brown:2010dt}\\
$|g_N\widetilde{J}^p_Z|$ & $< 6\times 10^{-58}$ &  $\widetilde{b}^p_Z$ & $< 7\times 10^{-29}$\;GeV & ~\cite{Peck:2012pt}\\[4pt]
$|g_N\widetilde{J}^n_X|$ & $< 7\times 10^{-63}$ & $\widetilde{b}^n_X$ & $(4.1\pm 4.7)\times 10^{-34}$\;GeV & ~\cite{Allmendinger:2013eya}\\
$|g_N\widetilde{J}^n_Y|$ & $< 7\times 10^{-63}$ &$\widetilde{b}^n_Y$ & $(2.9\pm 6.2)\times 10^{-34}$\;GeV & ~\cite{Allmendinger:2013eya}\\
$|g_N\widetilde{J}^n_Z|$ & $< 6\times 10^{-59}$ &  $\widetilde{b}^n_Z$ & $< 7\times 10^{-30}$\;GeV & ~\cite{Peck:2012pt}\\
\hline\hline
\end{tabular}
\caption{Constraints on  $|g_N\widetilde{J}_J^{N}|$ from co-magnetometer experiments, assuming the Sun as a source. The label $\perp$ indicates the equatorial ($X$-$Y$) plane, i.e., $\widetilde{b}^p_\perp = \sqrt{(\widetilde{b}^p_X)^2 + (\widetilde{b}^p_Y)^2 }$. The limits are valid for the mass of the scalar $m_\phi\lesssim 10^{-18}\,{\rm eV} $. }
\label{tab2}
\end{table}

\subsection{Potential peculiarities}
Let us note a couple of peculiarities of the measured quantity given in Eqs.~\eqref{VJpoint} and \eqref{a0}, \eqref{eq:m2} that can be calculated with Eq.~\eqref{eq:originalsourceint}. In contrast to the usual measurement in fifth force experiments, the measured quantity is not the force or some other derivative of the potential but effectively the potential itself. The reason is that the spin-dependent potential's energy depends on the spin's orientation with respect to the Lorentz-violating vector. When the experiment rotates relative to the fixed background, the energy changes by a size comparable to the effective potential. It acts like a fixed vectorial background field, akin to a magnetic field, as discussed in the previous section~\ref{sec:magnetic}. The size of the background field is determined by the source integral in Eq.~\eqref{eq:originalsourceint}
and features a Yukawa-like behaviour approaching $\sim 1/r$ in the massless limit. While the measured quantity in the end is a torque and, therefore, a force, its size is therefore determined by this much slower drop-off, which is different from the usual inverse square law. The first consequence is the incredible strength of the limits shown in Figure~\ref{fig:results}, which are much stronger than the already strong constraints on Lorentz-preserving coefficients.

However, the peculiarity of this result goes a bit deeper. To see this, we note that, in Eq.~\eqref{eq:originalsourceint}, all sources contribute, and the only decoupling of very distant sources occurs due to the mass $m_\phi$ of the scalar or equivalently, the range $\lambda=1/m_\phi$ of the potential. 
It therefore seems appropriate to also consider the effects of more distant sources. We could now consider the combined effect of the stars in our surroundings or even the galaxy\footnote{Indeed, taking account of the higher density in our local galactic environment the bound would be stronger by several orders of magnitude for masses up to the inverse of the size of the galactic disc $m\sim 10^{-27}\,{\rm eV}$.}. However, the main point that we want to make with the following discussion is that the potential features a very peculiar long distance behavior, resulting from far away matter. As a simple toy case we therefore immediately jump to consider the Universe as a whole, treated as having a homogeneous constant density of matter.

Let us start by approximating the Universe as homogeneous on very large distances. For the moment we neglect expansion. This should be appropriate for distances much smaller than the Hubble scale. We can then evaluate Eq.~\eqref{eq:originalsourceint} with a homogeneous density $\rho_{\rm{average}}$ and integrate in a spherically symmetric way up to a distance $R$\footnote{One may wonder whether our method of integrating over the source integral is indeed appropriate for the case of an essentially infinite homogeneous density. To justify this let us briefly consider the field equation of motion for a scalar coupling $g_B$ and a scalar number density $\rho/u$. Taking everything to be homogeneous this reads,
$\ddot{\phi}+3H\dot{\phi}+m^2_\phi\phi=-g_{B}\rho/u.$
Solving for the static case, we obtain
$\phi_{0}=-g_{B}\frac{\rho}{um^2_\phi}$.
Taking into account that in the source integral the coupling is factored out, this agrees with Eq.~\eqref{cr4}. Therefore, as long as the $m_\phi\gtrsim 1/R$ the source integral technique gives the correct result for the homogeneous case.},
\begin{equation}
    f(r)=- \frac{\rho_{\rm{average}}}{4\pi u}\int ^R_0 4\pi r dr e^{-\frac{r}{\lambda}}\,.
    \label{cr1}
\end{equation}
After evaluating the integral, we obtain
\begin{equation}
    f(R)=-\frac{\rho_{\rm{average}}}{u}\Big[\lambda^2(1-e^{-\frac{R}{\lambda}})-\lambda R e^{-\frac{R}{\lambda}}\Big].
    \label{cr2}
\end{equation}
Eq. \eqref{cr2} yields two limiting solutions by comparing the sizes of $\lambda$ and $R$.
\begin{enumerate}
    \item \emph{Long Range} $(\lambda\gg R).$ In this case, the range of the new force exceeds the size of the system, indicating that the mediator of the force is ultralight. As a result, the exponential term can be expanded, and Eq.~\eqref{cr2} simplifies to
    \begin{equation}
        f\approx-\frac{\rho R^2}{u}.
        \label{cr3}
    \end{equation}
   Note that, when taking $R\sim 1/H$ the expansion of the Universe cannot be neglected. We discuss this further below.
    \item \emph{Short Range} $(\lambda \ll R).$ In this case, the range of the new force is shorter than the size of the system, indicating that the mediator is massive compared to the ultralight case (case 1). Consequently, in the short-range limit, the exponential term cannot be expanded, and Eq.~\eqref{cr2} takes the following form
    \begin{equation}
       f\approx -\frac{\rho\lambda^2}{u}=-\frac{\rho}{u m^2_\phi}.
       \label{cr4}
    \end{equation}
\end{enumerate}

As we can see, the source integral diverges with the chosen distance $R$ in the massless limit. Accordingly, the constraints get tighter and tighter as we include increasingly distant sources\footnote{Due to the Lorentz-violating effect, there is also no cancellation of source contributions from the different directions in the centre of a homogeneously charged and isotropic source configuration.}. This rather strange behaviour is indicated by the purple line and region in Figure~\ref{fig:results} 
, where we have used a rough value $\rho=5\times 10^{-31}\,{\rm g}/{\rm cm}^3$ for the average baryon density in the Universe~\cite{Planck:2018vyg,uniwiki}. It suggests a non-decoupling of far distant objects.

As already mentioned for distances comparable to the Hubble scale the approximation of a static potential breaks down, and we should probably integrate no further than a distance of the order of the Hubble horizon before relativistic effects should be considered. 
Implementing this with a cutoff at a distance $\sim 1/H$ this is visible by the flattening of the curves at low masses.

For smaller or even vanishing mass we need to consider the time-dependent but still homogeneous problem. Let us consider the massless case\footnote{For non-vanishing mass $m_\phi\gg H$ the solution performs oscillations around the value of Eq.~\eqref{cr4}. However, these oscillations are exponentially damped by the expansion beyond the point $H(t)\lesssim m_\phi$ and we can therefore neglect them if the the mass is larger than today's Hubble scale. For masses less than today's Hubble scale the massless approximation should be a good indication of the behavior.}.Taking, for simplicity, the Hubble scale to be constant and starting from an initially vanishing field value and field velocity we obtain, from solving the field equation for spatially constant fields,
\begin{equation}
    \phi(t)=-g_{B}\frac{\rho}{u}\frac{\exp(-3Ht)-1+3Ht}{9H^2}\sim -g_{B}\frac{\rho}{u}\frac{1}{H^2}.
\end{equation}
Here, in the right hand side approximation we have assumed $Ht\sim 1$. Comparing with the source integral for the long-range case, Eq.~\eqref{cr3}, we can see that inserting $R\sim 1/H$ is indeed a decent approximation on the order of magnitude level. Indeed, we should take this as a caveat that the results in this range should be taken with a bit of caution and more as an indication of the order of magnitude. In particular, since our estimate also made simplistic assumptions on the initial conditions.

In any case, we
think that the long-range behaviour observed here indicates a potential sickness of the Lorentz-violating theory's long-distance infrared behaviour. Having noted this, we nevertheless proceed with our phenomenological analysis.

\section{Red Giant bounds on Lorentz violation}
\label{sec:redgiant}

Searches for long-range forces, as discussed in Sec.~\ref{sec:potentials}, typically have sensitivity at very low scalar masses, $m_{\phi}\ll 10^{-10}~{\rm eV}$. At higher scalar masses $m_\phi\lesssim \mathcal{O}(10)~\mathrm{keV}$, astrophysical systems such as RGs are often sensitive to extra energy losses caused by particle production~\cite{Raffelt:1996wa,Raffelt:2006cw}.

\begin{figure}
\centering
  \includegraphics[width=5cm]{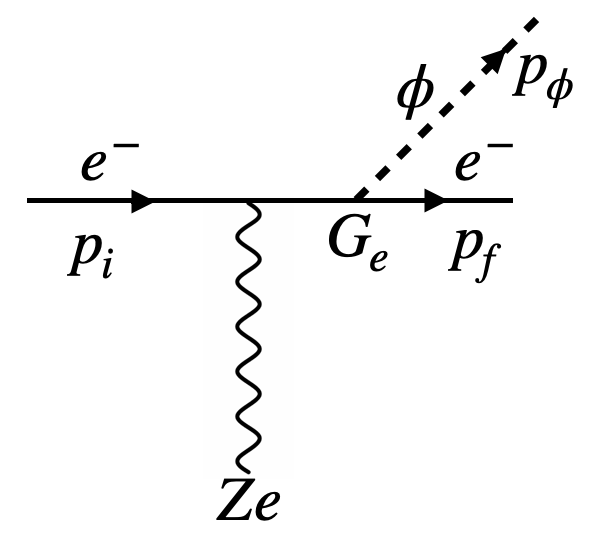}
  \caption{Scalar emission through a Lorentz-violating coupling in electron-ion bremsstrahlung process.}
  \label{fig:brem}
\end{figure}

In this section, we therefore consider the thermal production of a boson $\phi$ in stars interacting with electrons via the Lagrangian in Eq.~\eqref{YLV}, where $F=e$. In particular, we focus on the bremsstrahlung process $e^{-}+Ze\to e^{-}+Ze+\phi$~\cite{Raffelt:1994ry}, in which an electron $e$ scatters off an ion with charge $Ze$, emitting a boson $\phi$ in the final state, as shown in Figure~\ref{fig:brem} (see Ref.~\cite{Carenza:2021osu} for the case of axionic bremsstrahlung and Ref.~\cite{Bottaro:2023gep} for the scalar case). Since in this process, $\phi$ is emitted by an electron, with the ``spectator'' ion exchanging a photon with the incoming electron, the analysis in this section allows us to set new constraints only on Lorentz-violating interactions with electrons. Electron-ion bremsstrahlung is the dominant production channel for new scalars in RG stars. In this context, a subleading contribution to the $\phi$ production comes from the electron-electron bremsstrahlung $e^{-}+e^{-}\to e^{-}+e^{-}+\phi$, in which an electron would emit the particle $\phi$ after scattering off another electron. However, the latter process is suppressed due to the electron degeneracy in the RG core~\cite{Raffelt:1996wa}. Therefore, here we will use the electron-ion bremsstrahlung process in RG stars to probe $\phi$ interactions with electrons.

To constrain the energy loss via particle production, we use observations of RG stars. Let us briefly recall the main features relevant for this. Stars in the Main Sequence phase burn hydrogen in the core to produce energy and halt the gravitational collapse~\cite{Kippenhahn:2012qhp}. When there is no more hydrogen to burn, stars with a mass $M\lesssim2M_{\odot}$ will develop an inert helium core, which is too cold to ignite helium fusion, surrounded by a burning hydrogen shell. Such a star is called RG, characterized by expanding outer layers that increase the stellar luminosity because of the larger radius, even though the surface temperature decreases. While the hydrogen shell burns, the produced helium is continuously accumulated in the core, increasing its density and the electron degeneracy. Being a degenerate system, the core shrinks to increase the Fermi pressure, counteracting the collapse. However, in this process, the gravitational attraction on the core surface also increases, heating up this region. Together with the external layer, the inner core is heated up until it ignites the burning of helium, roughly at $\rho \approx 10^6~{\rm g}/{\rm cm}^3$ and $T\approx 10^{8}~{\rm K} = 8.6$~keV~\cite{Raffelt:1996wa,Kippenhahn:2012qhp}. This process, the so-called helium flash, is extremely fast because the helium burning rate is steeply dependent on temperature.
The properties of the RG when the helium flash happens determine the location of the so-called Red Giant Branch (RGB) tip in the colour-magnitude diagram~\cite{Kippenhahn:2012qhp,1970AcA....20...47P}. These properties are very sensitive to exotic energy emissions. As a matter of fact, exotic losses would delay the helium ignition, making the RGB tip brighter~\cite{Raffelt:1994ry,Caputo:2024oqc,Carenza:2024ehj}. Through a comparison with numerical simulations~\cite{Catelan:1995ba}, Ref.~\cite{Raffelt:2006cw} proposed to constrain novel particle emission by requiring that for a one-zone RG model with $T = 8.6$~keV and average density $\rho = 2\times 10^5$~g~cm$^{-3}$ the exotic energy loss should not exceed about $10$~erg~g$^{-1}$~s$^{-1}$~(see also \cite{Caputo:2024oqc} for a more recent discussion). We mention here that at these conditions, the standard neutrino emission is about 4~erg~g$^{-1}$~s$^{-1}$~\cite{Caputo:2024oqc}. Recently, constraints on the axion-electron coupling have been re-evaluated by inserting axions in stellar evolution codes and comparing the RGB tip obtained in the presence of axions with the observed ones. The most recent bound from observations of the RGB tip excludes $g_{ae}\lesssim 1.5~\times 10^{-13}$~\cite{Capozzi:2020cbu,Straniero:2020iyi}.

In this work, in order to estimate the RG bound on Lorentz-violating interactions, inspired by the strategy proposed in \cite{Raffelt:2006cw}, we evaluate the $\phi$ production via electron-ion bremsstrahlung using a one-zone model with $\rho = 2.5\times 10^{5}$~g~cm$^{-3}$ and temperature $T=8.6$~keV. A more refined analysis would require the inclusion of the $\phi$ production rates in stellar evolution codes, but this is beyond the scope of this work, especially since the simplified approach is compatible with numerical results~\cite{Catelan:1995ba,Raffelt:2006cw}.

\subsection{Production via electron-ion bremsstrahlung}
In the presence of Lorentz-violating interactions, the electron-ion bremsstrahlung matrix element is
\begin{equation}
    \mathcal{M}_{j}=\frac{\,Z_{j}e^{2}}{|\textbf{q}|(|\textbf{q}|^{2}+k_{S}^{2})^{1/2}}\\
    \times\bar{u}(p_{f})\left[G_e\,\frac{1}{\slashed{P}-m_{e}}\gamma^{0}+\gamma^{0}\frac{1}{\slashed{Q}-m_{e}}G_e\,\right]u(p_{i})\,,
\label{eq:matel}
\end{equation}
where $G_e$ is given by Eq.~\eqref{YLV}\footnote{Here, for simplicity we neglect the Lorentz-violating second-rank tensorial coupling $L_{\mu\nu}^e$ and focus on the vector and pseudovector coefficients.}, $u(p_{i}), u(p_{f})$ are the electron spinors, $p_{i}=(E_i,\textbf{p}_i)$ and $p_{f}=(E_f,\textbf{p}_f)$ are the four-momenta of the initial and final electrons, $P=p_{f}+p_\phi$, $Q=p_{i}-p_\phi$, and $\textbf{q}=\textbf{p}_{f}+\textbf{p}_{\phi}-\textbf{p}_{i}$  is the momentum transfer, with $p_\phi=(\omega_\phi,\textbf{p}_\phi)$ the four-momentum of the $\phi$ particle. The term $[|\textbf{q}|(|\textbf{q}|^{2}+k_{S}^{2})^{1/2}]^{-1}$ is the Coulomb propagator in a plasma and $k_{S}$ is the Debye screening scale given by \cite{Raffelt:1985nk}
\begin{equation}
k_S^2 = \frac{4\pi \alpha \sum_j Z_j^2 n_j}{T}\,,
\label{eq:ks}
\end{equation}
where $n_j$ is the number density of ions with charge $Z_j\,e$ and $\alpha$ is the fine structure constant. 

From the matrix element, we can compute the emission rate per unit volume and energy \cite{Carenza:2021osu}
\begin{equation}
    \frac{d^{2}n_{\phi}}{dt\,d\omega_{\phi}}=\int\frac{2d^{3}\textbf{p}_{i}}{(2\pi)^{3}2E_{i}}\frac{2d^{3}\textbf{p}_{f}}{(2\pi)^{3}2E_{f}}\frac{d\Omega_{\phi} \omega_\phi |\textbf{p}_\phi|}{(2\pi)^{3}2\omega_\phi}\\
    (2\pi)\delta(E_{i}-E_{f}-\omega_{\phi})\,|\overline{\mathcal{M}}|^{2}f_{i}(1-f_{f})\,,
\label{eq:dn}
\end{equation}
where $\Omega_\phi$ is the solid angle over which $\phi$ is emitted, $f_{i,f}$ are the electron distribution functions, the factors $2$ account for the electron spins, and $|\overline{\mathcal{M}}|^{2}=\frac{1}{4}\sum_{j} n_j \sum_{s} |\mathcal{M}_j|^{2}$ is the matrix element in Eq.~\eqref{eq:matel} averaged over the electron spins and summed over all the target ions. Notice that Eq.~\eqref{eq:dn} is obtained in the long-wavelength approximation, neglecting the momentum transferred to the scalar in the scattering, but considering it in the energy balance as shown by the Dirac delta. The explicit expression of $|\overline{\mathcal{M}}|^2$ depends on the considered interaction. Since its computation is non-trivial, we performed it with the help of the FeynCalc package~\cite{Kublbeck:1990xc,Shtabovenko:2016sxi,Shtabovenko:2020gxv}, obtaining the results shown in Appendix~\ref{app:Matrix} and available at \cite{github}.

In the absence of Lorentz-violating interactions, the integrals in Eq.~\eqref{eq:dn} can be reduced to a simpler form given in~\cite{Carenza:2021osu}.
However, this simplification cannot be done when Lorentz-violating terms are included since the matrix element would feature terms involving the scalar product between $I_\mu^e$, $J_\mu^e$ and the four-momenta $p_i$, $p_f$ and $p_\phi$.

Finally, the emissivity, i.e. the energy emitted per unit mass and time via $\phi$ emission is,
\begin{equation}
    \varepsilon_\phi=\frac{1}{\rho}\int_{m_\phi}^\infty d\omega_\phi\,\omega_\phi\,\frac{d^2n_\phi}{dtd\omega_\phi}\,.
\label{eq:emissivity}
\end{equation}

\subsection{Production in the Red Giant core}
The integrations in Eqs.~\eqref{eq:dn} and \eqref{eq:emissivity} involving Lorentz-violating vector and pseudovector coefficients can be done using a Monte Carlo simulation. To do so, 
we model the RG core as a He core with density $\rho = 2\times 10^{5}$~g~cm$^{-3}$ and temperature $T=8.6$~keV. This implies that the Debye screening scale is $k_S = 99.3$~keV, where in Eq.~\eqref{eq:ks} we used $Z^2\,n=Z(Z/A)\rho/m_u$, with $A=2Z=4$ and $m_u=931$~MeV.  

\subsubsection{Scalar and pseudoscalar interactions}
The emission via electron-ion bremsstrahlung induced by scalar and pseudoscalar interactions has already been considered in the literature; see, e.g. Refs.~\cite{Raffelt:1989zt,Raffelt:1996wa,Carenza:2021osu,Bottaro:2023gep}. Therefore, in order to check our simulations, we compared them against scalar and pseudoscalar cases.

If we consider only pseudoscalar interactions, $G_e=i\,g^{\prime}_e\,\gamma_5$, our analysis would correspond to the axion case, i.e $\phi=a$ and $g^{\prime}_e = g_{ae}$. The axion emission via electron bremsstrahlung was computed in Ref.~\cite{Carenza:2021osu} for any degree of degeneracy. For RG conditions, one finds
\begin{equation}
    \varepsilon_{\phi}=2.0\,{\rm erg}~{\rm g}^{-1}~{\rm s}^{-1} \left(\frac{g^{\prime}_e}{10^{-13}}\right)^{2}\left(\frac{T}{{\rm keV}}\right)^{2}\mathcal{F}_{\rm ps}\left(\eta,\frac{m_{\phi}}{T},\frac{m_{e}}{T},\frac{k_{S}}{T}\right)\,,
\label{eq:epsint_ps}
\end{equation}
where $\mathcal{F}_{\rm ps}$ is a complicated function which includes the integrals in Eq.~\eqref{eq:dn} and depends on the details of the $\phi$ interactions with electrons described by $|\overline{\mathcal{M}}|^2$. We stress that $\mathcal{F}_{\rm ps}$ is a function of both the axion mass and stellar plasma properties, including the electron degeneracy parameter $\eta=(\mu_e - m_e)/T=6.16$\cite{Carenza:2021osu}, with $m_e$ and $\mu_e$ the electron mass and chemical potential, respectively\footnote{The expression in Eq.~\eqref{eq:epsint_ps} is different from Eq.~(4) in Ref.~\cite{Carenza:2021osu}, due to different conditions. Here, we consider a helium core with $Z=2$, while in Ref.~\cite{Carenza:2021osu} $Z=1$ was considered for simplicity (see \cite{Carenza:2021osu} for more details). Therefore, the relation $k_s^2\,T = 4 \pi \alpha n$ was used, leading to a different expression.}. 

Analogously, for $G_e=\,g_e$ we obtain the scalar case, for which the emissivity is the same as in Eq.~\eqref{eq:epsint_ps}, but replacing $g^{\prime}_e \to g_e$ and $\mathcal{F}_{\rm ps}\to \mathcal{F}_{\rm s}$, a different function of the stellar properties taking into account the scalar interaction. The scalar emission via electron-nucleus bremsstrahlung was considered in Ref.~\cite{Bottaro:2023gep}. Under RG conditions and assuming $m_\phi \ll T$, we find for the pseudoscalar case $\mathcal{F}_{\rm ps}=9.3\times 10^{-3}$, while $\mathcal{F}_{\rm s}=23.0$. These values imply the ratio $\mathcal{F}_{ps}/\mathcal{F}_{s}\approx 4.0\times 10^{-4}$. This is in good agreement with the results in Ref.~\cite{Bottaro:2023gep}, showing that for degenerate electrons, the ratio between the pseudoscalar and scalar production rates is $\sim\pi^2/5 \times (T/m_e)^2$. For our fixed conditions in the RG cores, electrons are partially degenerate and $\pi^2/5 \times (T/m_e)^2 \approx 5.6\times 10^{-4}$, compatible with expectations. We numerically checked that the agreement improves for more degenerate conditions. Since electrons in RGs are partially degenerate, the integrals $\mathcal{F}_{\rm s}$ and  $\mathcal{F}_{\rm ps}$ need to be evaluated for different values of the temperature. For highly degenerate conditions $(\eta \gg 10)$ $\mathcal{F}_{\rm ps}\propto T^2 $, while $\mathcal{F}_{\rm s}$ is approximately independent of the temperature, implying that $\varepsilon_{\rm ps}\propto T^4$ for pseudoscalar and $\varepsilon_{\rm s}\propto T^2$ for scalar interactions, in accordance with literature (see, e.g., Ref.~\cite{Bottaro:2023gep} for scalar and Ref.~\cite{Raffelt:1989zt} for pseudoscalar particles).

\subsubsection{Vector and pseudovector interactions}

In the presence of the Lorentz-violating 4-vectors $I^e_\mu$ and $J^e_\mu$, we need to consider the complete integral in Eq.~\eqref{eq:dn}. Additionally, one has to take into account that the incoming electron is boosted in a given direction ${\bf v}=(v_{X},v_{Y},v_{Z})$ compared to the  $I^e_\mu$ and $J^e_{\mu}$ vectors. To calculate this effect, we perform a Monte Carlo by generating $10^4$ times $I^e_\mu$ and $J^e_\mu$, pointing in different directions, and for each case we compute the $I^e_\mu$ and $J^e_{\mu}$ vectors in the electron frame by transforming them with the appropriate Lorentz boost transformation, i.e. $I^e_{\mu}\to\Lambda^{\nu}_{\mu}I^e_{\nu}$ and $J^e_{\mu}\to\Lambda^{\nu}_{\mu}J^e_{\nu}$. We fit the obtained emissivity with generic bilinears in $I^e_0,\,I^e_i$ and $J^e_0,\,J^e_i$, where $i=X,Y,Z$, finding that the emissivity scales as $a_1 \, |I^e_0|^2 + a_2 \, |\textbf{I}^e|^2$ for vector interactions and as $b_1 \, |J^e_0|^2 + b_2 \, |\textbf{J}^e|^2$ for the pseudovector case, with $a_i$ and $b_i$ denote different coefficients. Both the emissivities are independent of mixed terms as $I^e_0\,I^e_i$ and $I^e_i\,I^e_j$, since each contribution from an electron coming from a given direction is balanced by an electron coming from the opposite one. Also, in this case, the emissivity can be written as
\begin{equation}
    \varepsilon_{\phi}=2.0\times 10^{26}\,{\rm erg}~{\rm g}^{-1}~{\rm s}^{-1} \left(\frac{T}{{\rm keV}}\right)^{2}\mathcal{F}_{\rm v}\left(|I^e_0|^2 , |\textbf{I}^e|^2,\eta,\frac{m_{\phi}}{T},\frac{m_{e}}{T},\frac{k_{S}}{T}\right)\,.
\label{eq:epsint_v}
\end{equation}
The emissivity for the pseudovector interaction can be obtained by simply replacing $\mathcal{F}_{\rm v}\to \mathcal{F}_{\rm pv}\left(|J^e_0|^2 , |\textbf{J}^e|^2,\eta,\frac{m_{\phi}}{T},\frac{m_{e}}{T},\frac{k_{S}}{T}\right)$. Again, since electrons in RGs are partially degenerate, one needs to evaluate the functions $\mathcal{F}_{\rm v}$ and $\mathcal{F}_{\rm pv}$ for the different temperatures. For highly degenerate electrons, both $\mathcal{F}_{\rm v}$ and $\mathcal{F}_{\rm pv}$ are independent of the temperature, implying that the emissivity scales as $T^2$ for both the vector and the pseudovector cases.
 
If we assume that both the Lorentz-preserving ($g_e$ and $g^{\prime}_e$) and the Lorentz-violating ($I^e_\mu$ and $J^e_\mu$) coefficients are non-vanishing, there will be some non-zero interference terms contributing to the emissivity. We find that only parity-even interference terms are non-vaninshing, i.e. scalar-vector and pseudoscalar-pseudovector. In this case, for typical RG conditions, we numerically find

\begin{eqnarray}
\label{eq:epsgen}
    \varepsilon_{\phi} &\approx&   1.5 \times 10^{28}\,{\rm erg}~{\rm g}^{-1}~{\rm s}^{-1}
    \\\nonumber
    &&\qquad\qquad\times
    \bigg[23.0\,{g_e^2} \left(\frac{T}{8.6~{\rm keV}}\right)^2+ 9.3\times 10^{-3}\,{g^{\prime\,2}_e} \left(\frac{T}{8.6~{\rm keV}}\right)^4 
    \\\nonumber
    &&\qquad\qquad\qquad+ 
    \bigg(56\, {I^e_0}^2 + 34\, {|\textbf{I}^e|}^2\bigg)\left(\frac{T}{8.6~{\rm keV}}\right)^2 + 
    \bigg(97 {J^e_0}^2 + 38{|\textbf{J}^e|}^2\bigg)\left(\frac{T}{8.6~{\rm keV}}\right)^2 
    \\\nonumber
    &&\qquad\qquad\qquad\qquad\qquad\qquad\qquad\qquad+ 
    59\, g_e\,I^e_0 \left(\frac{T}{8.6~{\rm keV}}\right)^2 + 0.87\,g^{\prime}_e\,J^e_0 \, \left(\frac{T}{8.6~{\rm keV}}\right)^3\bigg]\,,
\end{eqnarray}
where the temperature dependence is obtained by slightly changing the value of $T$, keeping the density and the other quantities fixed. This relation shows that the pseudoscalar emission (and its interference term with the pseudovector one) is suppressed compared to the other terms. This is in agreement with the non-relativistic potential shown in Eq.~\eqref{pots}. There, all the terms, except the pseudoscalar one, feature leading terms behaving as $r^{-1}$. In contrast the pseudoscalar-induced potential decreases as $r^{-2}$ or faster, cf.~Eq.~\eqref{potss}. Since electrons in the RG core are degenerate and non-relativistic, further physical intuition can be gained by looking at the degenerate and non-relativistic limit ($E_F\gtrsim m_e \gg T \sim \omega_\phi$, with $E_F$ the Fermi energy) of the matrix elements shown in Appendix~\ref{app:Matrix}. Indeed, in these conditions the electron momenta are close to the Fermi surface $|\textbf{p}_i|=|\textbf{p}_f|= k_F$, with $k_F=\beta_F E_F$, $\beta_F$ being the Fermi velocity. As further discussed in Appendix~\ref{app:Matrix}, in such a limit the integrals over the energies to compute the emissivity are analytical, leading to $\varepsilon_\phi\propto T^2$ for all the couplings but for the pseudoscalar ($\varepsilon_\phi\propto T^4$) and the pseudoscalar-pseudovector interference term  ($\varepsilon_\phi\propto T^3$).

\subsection{Bounds from Red Giants}
The most recent RG bound on axions excludes $g^{\prime}_e \gtrsim 1.5\times 10^{-13}$~\cite{Capozzi:2020cbu,Straniero:2020iyi}. Therefore, one can set a bound on the dimensionless coefficients by requiring that the emissivity for the fixed RG condition is equal to the axion case for $g^{\prime}_e = 1.5\times 10^{-13}$. For typical RG conditions, the axion emissivity can be written as
\begin{equation}
    \varepsilon_a =\left(\frac{g^{\prime}_e }{10^{-13}}\right)^{2}1.45~{\rm erg}~{\rm g}^{-1}~{\rm s}^{-1}\,,
\label{eq:epsilona}
\end{equation}
implying $\varepsilon_a \lesssim 3.3$~erg g$^{-1}$ s$^{-1}$ at the bound~\cite{Capozzi:2020cbu,Straniero:2020iyi}\footnote{Notice that the requirement $\varepsilon_a \lesssim 10$~erg~g$^{-1}$~s$^{-1}$ would lead to the constraint $g^{\prime}_e \lesssim 2.6\times 10^{-13}$, in agreement with \cite{Raffelt:2006cw}.}.  
By requiring $\varepsilon_\phi < 3.3$~erg~g$^{-1}$~s$^{-1}$, one obtains a constraint on the general combination of the couplings valid for $m_\phi \ll \mathcal{O}(10)$~keV

\begin{eqnarray}
\label{eq:genbound}
\big( 23\,g_e^2 \!\!&+&\!\! 9.3\times 10^{-3}\,{g^{\prime\,2}_e} + 56\, {I^e_0}^2 + 34\, {|\textbf{I}^e|}^2  \\\nonumber
   && \qquad\qquad +97 {J^e_0}^2 + 38 {|\textbf{J}^e|}^2 + 59\, g_e\,I^e_0 + 0.87\,g^{\prime}_e\,J^e_0 \, \big) < 2.2\times 10^{-28}\,.
\end{eqnarray}
In Table~\ref{tab:RGlimits}, we report the values of the individual bounds for the considered couplings, obtained by setting all the other interactions to zero in Eq.~\eqref{eq:genbound}. These constraints would be slightly strengthened by a few per cent if one includes the electron-electron bremsstrahlung, suppressed due to the electron degeneracy in the RG core~\cite{Raffelt:1996wa}. In particular, for the scalar interaction, we find a constraint $g_e\lesssim 3\times 10^{-15}$, in close agreement with the result obtained in Ref.~\cite{Bottaro:2023gep} by rescaling axion bounds from RGs. Here, it is worth mentioning that recent studies found stronger constraints also on the scalar interactions, excluding $g_{\phi e}\gtrsim 0.7\times 10^{-15}$ from resonant plasmon conversion in RGs~\cite{Hardy:2016kme} and $g_{\phi e}\gtrsim 0.4\times 10^{-15}$ using the white-dwarf luminosity function~\cite{Bottaro:2023gep}. Therefore, it is plausible that astrophysical observations may lead to slightly stronger constraints (by factors of a few) on the vector and pseudovector couplings through a more refined analysis of the production in RGs or using a different observable. Since we find that the temperature dependence of the vector and pseudovector emissivities is the same as the scalar one ($\varepsilon_\phi \propto T^2$), we expect that the strongest bounds on these couplings would come from the white-dwarf luminosity function~\cite{Bottaro:2023gep}, constraining the temporal and the spatial components of $I^e_\mu$ and $J^e_\mu$ to be $\lesssim {\rm few}\,\times 10^{-16}$. This result would not affect our conclusions. Indeed, our goal here is to show that astrophysical observations offer the possibility to constrain different combinations of the Lorentz-violating couplings, probing masses several orders of magnitude larger than the ranges discussed in the previous sections. A comprehensive analysis of astrophysical constraints is beyond the scope of this paper. Therefore, here we focus solely on the RG bound.

\renewcommand{\arraystretch}{1.25}
\begin{table}[t!]
\centering
\begin{tabular}{c c}
\hline\hline
Parameter 
& Constraint  
\\
\hline
$g_e$  & $3.1 \times 10^{-15}$ \\
$g^{\prime}_e$  &  $1.5\times 10^{-13}$\\
$I^e_0$ & $2.0\times 10^{-15}$\\
$|{\bf I}^e|$ & $2.5\times 10^{-15}$\\
$J^e_0$ & $1.5\times 10^{-15}$\\
$|{\bf J}^e|$ & $2.4\times 10^{-15}$\\
\hline\hline
\end{tabular}
\caption{RG constraints on the scalar and pseudoscalar couplings $g_e$ and $g^{\prime}_e$, as well as on the strength of the temporal and spatial components of the Lorentz-violating vector $I^e_\mu =(I^e_0,{\bf I}^e)$ and pseudovector $J^e_\mu =(J^e_0,{\bf J}^e)$ coefficients. Here we assume that $m_\phi \ll \mathcal{O}(10)$~keV. Each limit is obtained by setting all the other couplings to zero in Eq.~\eqref{eq:genbound}.}
\label{tab:RGlimits}
\end{table}

We conclude this section with a comparison with the constraints on $|g_S\,\widetilde{J}^e_X|$ shown in Figure~\ref{fig:results}\footnote{Let us reiterate that, strictly speaking, the results from constraints on long-range potentials shown in Figure~\ref{fig:results} are for $g_S=g_N$, however limits on an electron coupling would have a similar order of magnitude.}. Neglecting the tensorial coupling, from Eq.~\eqref{tildebasis} we obtain $\widetilde{J}^e_J=J^e_J$, $J=X,\,Y,\,Z$. Therefore, the bound on $g_e\widetilde{J}_J^e$ in RGs is obtained from the condition
\begin{equation}
    (23\,g_e^2 + 38 |\textbf{J}^e|^2)<2.2\times10^{-28}\,.
\end{equation}
Due to the isotropy of the scattering, the direction of $\textbf{J}^e$ is not relevant in this case. Therefore, to make a comparison with constraints in Figure~\ref{fig:results}, we assume $\textbf{J}^e$ pointing in the $X$-direction in the RG rest frame. This leads to the constraint on the product\footnote{The inequality $a\,x^2 + b\,y^2 < c$, with $a,b,c$ constants and $x,\,y$ variables, implies $x\,y<\frac{c}{2\sqrt{ab}}$.} $g_e\,\widetilde{J}^e_X\equiv g_e\,J^e_X<3.7\times10^{-30}$ for $m_\phi \lesssim 8$~keV. Note that, as our calculation is done for massless scalars, its validity is restricted to values less than the temperature. This represents our strongest constraint on $g_e\,\widetilde{J}^e_J$ in the $10^{-10}~{\rm eV}\lesssim m_\phi \lesssim 8$~keV range and is shown in red in Figure~\ref{fig:results}.

The RG bound is depicted as a dashed line for $m_{\phi}\lesssim 10^{-9}$~eV, since in this regime the scalar may introduce new long-distance interactions that could alter the RG stellar structure. In such a situation, the analysis would need to be adjusted to account for potential modifications to the stellar equilibrium. This would require the self-consistent inclusion of the low-mass boson in stellar evolutionary codes, which is beyond the scope of this work. As a back-of-the-envelope estimation, we assume that the new force mediated by the boson affects the stellar structure when the standard pressure in the center balancing gravity becomes smaller than the pressure due to the Yukawa force. For a simple estimate we use, for the ordinary pressure, the electron degeneracy pressure $P_e=(3\pi^2)^{2/3}(n_{e})^{5/3}/(5m_{e})$~\cite{1939isss.book.....C}, where $n_{e}$ is the electron density in the core. For the one caused by the Yukawa force, we adapt the formula for the gravitational pressure inside a homogeneous sphere to a force of range $\lambda$. The gravitational pressure is given by $ P_G = 3/(8\pi)\, (GM/R^2) \, (M/R^2)=(2\pi/3)\,GR^2\rho^2$~\cite{1939isss.book.....C}, where $\rho$ is the density.
For the Yukawa force we then use $g_{e}\sim J_{e}$, and replace $R\to\lambda\sim m^{-1}$ and $Gm_{N}^2\to g^{2}_{e}/(4\pi)$, leading to $P_G\to P_\phi = g_e^2\,\lambda^2\,\rho^2/6 m_N^2 $. For the densities we use values from the center of the RG. Generically, we find that the Yukawa pressure exceeds the degeneracy pressure when $m_\phi\lesssim 4\times 10^{-10}~{\rm eV}\, (g_e/3\times10^{-15})$. Thus, for values of the coupling close to our RG limit, $P_e>P_G$ for $\lambda\gtrsim \mathcal{O}(\rm 100\, m)$, i.e. $m_\phi\lesssim 10^{-9}$~eV. Hence, in this region the scalar would affect gravity and the limit would strictly speaking not be self-consistent.

\section{Discussions, Conclusions and Outlook}\label{sec:conclusions}

In this paper, we have obtained constraints on the dimensionless Lorentz-violating couplings in a setup where LV occurs via a relatively light new particle coupled to SM fermions that extends the SME~\cite{Colladay:1996iz, Colladay:1998fq,Kostelecky:2003fs} by an additional degree of freedom. 
This allows for long-range forces and real particle production in addition to Lorentz symmetry violation.
Notably, at very low masses $\ll 10^{-10}\,{\rm eV}$ the strongest constraints arise from probing spin-dependent long-range forces in solar system scale, whereas at higher masses up to $\sim 10\,{\rm keV}$ astrophysical (RG) constraints dominate.

More concretely, we have considered an ultralight scalar coupling to polarized electrons or nucleons in Earth-based experiments and to unpolarized electrons or nucleons in the Sun and Earth through Lorentz-violating interactions, mediating a long-range force.
Using torsion-balance and magnetometer experiments, which are sensitive to such forces, we place constraints on Lorentz-violating couplings, particularly on a specific combination, $ g_S \widetilde{J}_J^{e,N}$, where $g_S$ denotes the monopole coupling of the scalar with the nucleons or electrons in the source and $\widetilde{J}_J^{e,N}$ denotes a combination of pseudovector and rank two tensor couplings with direction $J=X,Y,Z$ with the electron/nucleon spin. The obtained limits $g_S \widetilde{J}_J^e\sim 10^{-60}$ are extremely strong, benefitting from the long-range nature of the interactions as well as the tremendous precision of the experiments.

Constraining the combination $ g_S \widetilde{J}_J^{e,N}$ is, of course, only a first step. Eq.~\eqref{YLV} and consequently also Eqs.~\eqref{pots} and \eqref{potss} contain a multitude of other Lorentz-violating coefficients. We think that many of them would also be amenable to being probed with suitable experiments similar to the ones considered in this paper.
For example, the $\widetilde{L}_j$ coefficients generate a product of sidereal and (suppressed) annual variations. To the best of our knowledge, no experiments have probed this product of variations. However, it seems plausible that an E{\"o}t-Wash type re-analysis, that jointly accounts for the Earth's axial rotation and revolution about the Sun, could provide first constraints on these coefficients.
Moreover, we have focused on constraining a velocity-independent, scalar-mediated Lorentz-violating potential generated by an unpolarized source.
Other Lorentz-violating interactions, which include additional velocity suppression or require polarization of the source, are not considered here but can likely be constrained using similar methods.

Another important feature of including a new light particle is that it can also be produced as a real final state in interactions between SM particles. This opens a different route for probing this scenario. Following the established fact that stars are an excellent tool to search for very feeble interactions of reasonably light particles~\cite{Raffelt:1996wa,Raffelt:2006cw,Caputo:2024oqc,Carenza:2024ehj} 
we also determine bounds on various Lorentz-violating couplings through the radiation of scalars produced in the electron-ion bremsstrahlung process, which contributes to the emissivity of RG stars. 
This allows us to set limits on the square of nearly\footnote{The tensorial coefficients should be amenable to being constrained with the same technique. However, the computation of the relevant matrix elements is more challenging and beyond the scope of this initial analysis. } all individual Lorentz-violating and non-violating couplings to electrons at the $\sim 10^{-30}$ level. 
Explicitly, this also constrains the combination
$ g_e \widetilde{J}_J^{e}\leq 3.7\times 10^{-30}$. 
These limits hold for scalar masses $m_\phi \lesssim \mathcal{O}(10)~\mathrm{keV}$, as determined by the temperature of the RGs, allowing us to constrain a mass range significantly larger than the other probes discussed in this work.
A further increase in the mass reach may be achieved through beam-dump experiments or collider searches, albeit at the price of requiring somewhat less feeble interactions. 

Finally, we would like to note that generally feebly interacting light bosons can also be constrained using black hole (BH) spin measurements through superradiance~\cite{Arvanitaki:2009fg,Arvanitaki:2010sy,Arvanitaki:2014wva,Stott:2018opm,Baryakhtar:2020gao,Mehta:2021pwf,Hoof:2024quk,Lambiase:2025twn}. 
Currently, these constraints apply to narrow regions where the inverse boson mass is comparable to the BH size. From the observed populations these are in two regions,  $10^{-13} \, {\rm eV} \lesssim m_\phi \lesssim10^{-11} \, {\rm eV}$~\cite{Arvanitaki:2014wva} and $10^{-20} \, {\rm eV} \lesssim m_\phi \lesssim 10^{-17} \, {\rm eV}$~\cite{Brito:2014wla,Stott:2018opm}. 
However, these constraints apply only to bosons with sufficiently weak self-interactions. To avoid an assumption about them we do not include these constraints in Fig.~\ref{fig:results}.

For our phenomenological purposes the mass of interest for the scalar is simply determined by the different physical conditions relevant to each experimental setup. For instance, by the distance between the source and the detector in torsion-balance and magnetometer-based experiments, and by the temperature of the thermal medium in RGs for stellar cooling bounds. While we do not delve into the mass-generation mechanism, we note that such a scalar can acquire mass through several theoretical frameworks. These include Planck-suppressed operators arising in quantum gravity or string theory \cite{Hui:2016ltb,Hubisz:2024hyz,Banerjee:2022wzk}, the clockwork mechanism \cite{Giudice:2016yja,Wood:2023lis}, or via mixing with the SM Higgs \cite{Batell:2022qvr}. The Higgs portal can, in principle, generate a mass for the scalar. However, achieving an ultralight mass through this Higgs vev necessitates an extremely small scalar-Higgs coupling, which in turn requires significant fine-tuning.

The bounds from torsion balance and magnetometer-based experiments are derived from the potential calculated using a tree-level Feynman diagram, where one vertex involves a scalar coupling and the other involves a Lorentz-violating coupling—specifically, a combination of pseudovector and rank-2 tensor operators. The scalar mediator couples to a macroscopic unpolarized source through a scalar interaction at one vertex, while at the other vertex, it interacts with polarized electrons or nucleons via a Lorentz-violating coupling. Importantly, the potential and hence the experimental sensitivity arises from this specific combination of couplings. Attempting to constrain the individual couplings separately and then multiplying them would correspond to a completely different physical setup, with distinct Feynman diagrams and resulting potentials. Therefore, such an approach would not yield the correct physical interpretation. Although, one could look at \cite{Cong:2024qly,OHare:2020wah} for the current constraints on the individual couplings.

Finally, at a more theoretical level, it would be interesting to understand better the causes and effects of several peculiarities associated with the Lorentz-violating scalar interactions. Notably, the measurable effects in potentials decrease with distance considerably slower than in the case of non-Lorentz-violating interactions. This holds in particular for spin-dependent interactions. Similarly, real particle production in stars also does not seem to exhibit the significant velocity suppression that is often associated with spin-dependent interactions. 

All in all, this shows that going beyond the SME and including new light degrees of freedom may be an interesting and powerful road to probing Lorentz violation.

\section*{Acknowledgements}
We would like to thank Rick Gupta for collaboration and discussions in the early stage of this project. GL and PC are grateful to Salvatore Bottaro and Arturo de Giorgi for helpful discussions. The authors would also like to thank the anonymous referee for fruitful comments and suggestions.
This article is based on the work from COST Actions COSMIC WISPers CA21106 and BridgeQG CA23130, supported by COST (European Cooperation in Science and Technology). PC is supported by the Swedish Research Council under contract 2022-04283. JJ acknowledges the EU for support via ITN HIDDEN (No 860881). GL acknowledges support from the U.S. Department of Energy under contract number DE-AC02-76SF00515 and, when this work was started, the EU for support via ITN HIDDEN (No 860881). NS acknowledges support from the U.K. Science and Technology Facilities Council under grants ST/T006048/1 and ST/Y004418/1, by the Deutsche
Forschungsgemeinschaft under the Heinz Maier Leibnitz Prize BeyondSM HML-537662082,
and by the Indiana University Center for Spacetime Symmetries. TKP is partially supported by the Italian Istituto Nazionale di Fisica Nucleare (INFN) through the ``QGSKY" project.

\appendix
\section{Lorentz-violating scalar-pseudovector coupled potential mediated by an ultralight scalar boson}\label{app1}
This appendix explicitly derives the Lorentz-violating scalar-pseudovector ($ \sim gJ$) coupling terms in the potential. The same procedure yields all other terms which have already been calculated in~\cite{Altschul:2012xu}.

\begin{figure}
\centering
  \includegraphics[width=7.5cm]{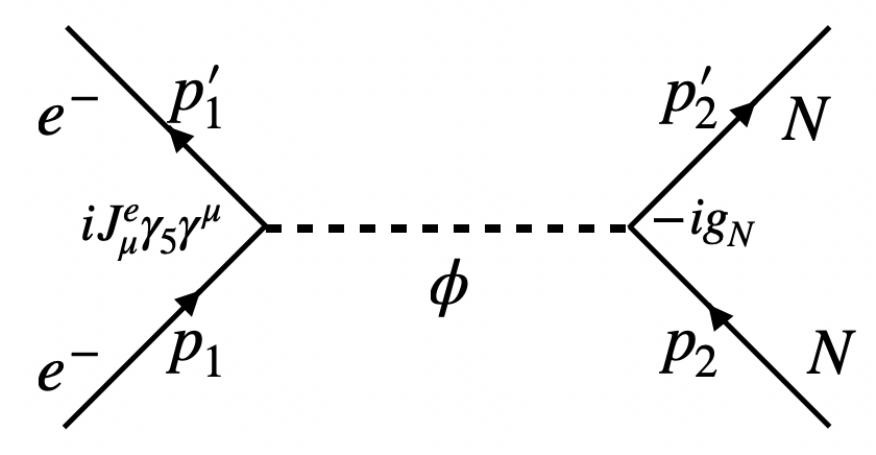}
  \caption{Scalar-pseudovector coupled Lorentz-violating potential mediated by an ultralight scalar boson}
  \label{newplota1}
\end{figure}

The amplitude of the electron-nucleon interaction mediated by an ultralight scalar, where the electron exhibits a Lorentz-violating pseudovector coupling and the nucleon interacts via a scalar coupling, is shown in Figure~\ref{newplota1}. It can be expressed as
\begin{eqnarray}
i\mathcal{M}&=&\bar{u}_{s^\prime_1}(p^\prime_1)(iJ^e_\mu\gamma_5\gamma^\mu)u_{s_1}(p_1)\frac{i}{q^2-m^2_\phi}\bar{u}_{s^\prime_2}(-ig_N)u_{s_2}(p_2)\nonumber\\
&=& \frac{ig_NJ^e_\mu}{q^2-m^2_\phi}\bar{u}_{s^\prime_1}(p^\prime_1)\gamma^\mu\gamma_5 u_{s_1}(p_1)\bar{u}_{s^\prime_2}(p^\prime_2)u_{s_2}(p_2),
\label{bk1}
\end{eqnarray}
where the momentum transfer is $q=p_1-p^\prime_1=p^\prime_2-p_2$ and $s_1$, $s_1^\prime$, $s_2$, $s_2^\prime$ refer the spin of each fermion. 

We work in the non-relativistic regime, where the components of the three-momentum for both electrons and nucleons are much smaller than their respective masses $(m_i)$. As a result, the particle's energy can be approximated as $E_i \approx m_i$. We choose the normalization condition $u^\dagger_{s^\prime}(p)u_s(p)=\delta_{ss^\prime}$ and write the expression of the positive energy spinor in the NR limit as
\begin{equation}
u_s(p)=\Big(1-\frac{\gamma_ip_i}{2m}\Big)\xi_s +\mathcal{O}(p^2) \,.
\label{bk2}
\end{equation}
Here, $\xi_s$ is the normalized eigenvector which satisfies $\xi_s^\dagger\gamma^0=\xi^\dagger_s$, $\gamma_0\xi_s=\xi_s$, and $\gamma_i$ denotes the Dirac gamma matrices, where $i$ runs from $1$ to $3$. Therefore, in the NR limit, we can write the Dirac bilinears using Eq.~\eqref{bk2} as
\begin{equation}
\bar{u}_{s^\prime}(p)u_{s}(p)=\delta_{s^\prime s}, ~J^e_0 \bar{u}_{s^\prime}(p)\gamma^0\gamma_5 u_s(p)=J^e_0\xi^\dagger_{s^\prime}\gamma_5\xi_s=0, ~J^e_i\bar{u}_{s^\prime}(p)\gamma^i\gamma_5 u_s(p)=-\left(\boldsymbol{\sigma}_e\cdot\boldsymbol{J}^{e}\right),
\label{bk3}
\end{equation}
where in the last expression, $s=s^\prime$. 

Evaluating Eq.~\eqref{bk1} in the non-relativistic limit, and with ${q^0}^2\ll |\mathbf{q}^2|$, we have,
\begin{equation}
i\mathcal{M}=-\frac{ig_N}{q^2-m^2_\phi}\left(\boldsymbol{\sigma}_e\cdot\boldsymbol{J}^{e}\right)\approx \frac{ig_N}{|\mathbf{q}|^2+m_\phi^2}\left(\boldsymbol{\sigma}_e\cdot\boldsymbol{J}^{e}\right)\,.
\label{bk4}
\end{equation}
With this, the potential reads
\begin{equation}
V(r)=-\int \frac{d^3 q}{(2\pi)^3}e^{i\mathbf{q}\cdot \mathbf{r}}\frac{g_N}{|\mathbf{q}|^2+m_\phi^2}\left(\boldsymbol{\sigma}_e\cdot\boldsymbol{J}^{e}\right)=-g_N \left(\boldsymbol{\sigma}_e\cdot\boldsymbol{J}^{e}\right)\Big(\frac{e^{-m_\phi r}}{4\pi r}\Big)=\left(\boldsymbol{\sigma}_e\cdot\boldsymbol{J}^{e}\right)g_N f(r),
\label{bk5}
\end{equation}
where $f(r)=-e^{-m_\phi r}/(4\pi r)$.

To find the potential associated with $J_0$, we need to consider a higher order term, $\mathcal{O}(p)$ of the spinor equation (Eq.~\eqref{bk2}). Thus, we can write
\begin{equation}
\bar{u}_{s^\prime}(p^\prime)\gamma^0\gamma_5 u_s(p)=\frac{1}{2m}\xi^\dagger_{s^\prime}(\gamma_i\gamma_5p^\prime_i+\gamma_j\gamma_5p_j)\xi_s=\frac{1}{2}(\sigma_i v^\prime_i+\sigma_jv_j)=\left(\boldsymbol{\sigma}_e\cdot\boldsymbol{v}^{e}_\mathrm{av}\right),
\label{bk6}
\end{equation}
where $s=s^\prime$ and $v^e_{\mathrm{av}}=(1/2)(v^e+{v^\prime}^e)$ is the average velocity. Therefore, we can write Eq.~\eqref{bk1} as
\begin{equation}
i\mathcal{M}=\frac{ig_NJ^e_0}{q^2-m^2_\phi}\left(\boldsymbol{\sigma}_e\cdot\boldsymbol{v}^{e}_\mathrm{av}\right).
\label{bk7}
\end{equation}
Thus, the non-relativistic potential becomes
\begin{equation}
V(r)=\int \frac{d^3 q}{(2\pi)^3}e^{i\mathbf{q}\cdot \mathbf{r}} \frac{g_N J^e_0}{|\mathbf{q}^2|+m_\phi^2} \left(\boldsymbol{\sigma}_e\cdot\boldsymbol{v}^{e}_\mathrm{av}\right)=-\left(\boldsymbol{\sigma}_e\cdot\boldsymbol{v}^{e}_\mathrm{av}\right)g_N J^e_0 f(r).
\label{bk8}
\end{equation}
 Eqs.~\eqref{bk5} and \eqref{bk8} denote the Lorentz-violating scalar-pseudovector coupled potential mediated by an ultralight scalar boson and the form of the potentials match to that of Eq.~\eqref{pots} that was obtained in~\cite{Altschul:2012xu}. Using a similar approach, the other possible potential functions, as presented in Eq.~\eqref{pots} and found in~\cite{Altschul:2012xu}, can also be derived.

\section{Spin averaged amplitude squared for the emission of Lorentz-violating scalar in electron-ion bremsstrahlung process}
\label{app:Matrix}

The emission of a Lorentz-violating scalar $\phi$ via electron-ion bremsstrahlung is represented by the diagram in Figure~\ref{fig:brem}. Assuming an ion of charge $Ze$, the matrix element can be written as
\begin{equation}
    \mathcal{M}=\frac{\,Ze^{2}}{|\textbf{q}|(|\textbf{q}|^{2}+k_{S}^{2})^{1/2}}\\
    \times\bar{u}(p_{f})\left[G_e\,\frac{1}{\slashed{P}-m_{e}}\gamma^{0}+\gamma^{0}\frac{1}{\slashed{Q}-m_{e}}G_e\,\right]u(p_{i})\,.
\end{equation}
In the above equation, $G_e$ includes Lorentz-violating coefficients and it is given by Eq.~\eqref{YLV}, $u(p_{i}), u(p_{f})$ are the electron spinors, $p_{i}=(E_i,\textbf{p}_i)$ and $p_{f}=(E_f,\textbf{p}_f)$ are the four-momenta of the initial and final electrons with mass $m_e$, $P=p_{f}+p_\phi$, $Q=p_{i}-p_\phi$, and $\textbf{q}=\textbf{p}_{f}+\textbf{p}_{\phi}-\textbf{p}_{i}$  is the momentum transfer, with $p_\phi=(\omega_\phi,\textbf{p}_\phi)$ the four-momentum of the $\phi$ particle with mass $m_\phi$. The term $[|\textbf{q}|(|\textbf{q}|^{2}+k_{S}^{2})^{1/2}]^{-1}$ is the Coulomb propagator in a plasma and $k_{S}$ is the Debye screening scale given by \cite{Raffelt:1985nk}
\begin{equation}
k_S^2 = \frac{4\pi \alpha \sum_j Z_j^2 n_j}{T}\,,
\end{equation}
where $n_j$ is the number density of ions with charge $Z_j e$ and $\alpha$ is the fine structure constant. From the matrix element, one can compute the spin averaged amplitude squared as 
\begin{equation}
   |\overline{\mathcal{M}}|^2=\frac{1}{4}\sum_j n_j \sum_s |{\mathcal{M}}_j|^2=\sum_j n_j \frac{Z_j^2\,e^4}{\mathbf{q}^2(|\mathbf{q}|^2+k_s^2)}\frac{1}{4}\sum_s |M|^2\,,
\end{equation}
where we sum over target species $j$ and average over spins $s$. In the helium core of RGs, we can write 
\begin{equation}
    |\overline{\mathcal{M}}|^2=\frac{n_{\rm He}\,Z_{\rm He}^2\,16\pi^2 \alpha^2}{\mathbf{q}^2(|\mathbf{q}|^2+k_s^2)}\frac{1}{4}\sum_s|M|^2\,,
\label{eq:Mbar}
\end{equation}
where $\frac{1}{4}\sum_s|M|^2$ includes the details of the considered interactions. Assuming that both the scalar $g^e$ and pseudoscalar $g^{\prime e}$ Lorentz-preserving coefficients, as well the vector $I^e_\mu\equiv I^e=(I^e_0,\textbf{I}^e)$ and pseudovector $J^e_\mu\equiv J^e=(J^e_0,\textbf{J}^e)$ Lorentz-violating coefficients are non vanishing, we have,
\begin{equation}
    \frac{1}{4}\sum_s|M|^2 = \mathcal{T}_{\rm s} + \mathcal{T}_{\rm ps} + \mathcal{T}_{\rm v} + \mathcal{T}_{\rm pv} + \mathcal{T}_{\rm s-v} + \mathcal{T}_{\rm ps-pv}\,,
\end{equation}
where $\mathcal{T}_{\rm s}$, $\mathcal{T}_{\rm ps}$, $\mathcal{T}_{\rm v}$ and $\mathcal{T}_{\rm pv}$ include only scalar, pseudoscalar, vector and pseudovector interactions, respectively, while $\mathcal{T}_{\rm s-v}$ and $\mathcal{T}_{\rm ps-pv}$ are scalar-vector and pseudoscalar-pseudovector interference terms.

Explicitly, in the limit of vanishing $m_\phi$ this results in the following expressions: 
\begin{equation}
\begin{aligned} 
\mathcal{T}_s&=g_e^2\times\bigg[\frac{m_e^2 \left(-p_\phi \cdot p_i - p_f \cdot p_i + 2 E_i (\omega_\phi + E_f) + m_e^2\right)}{\left(p_\phi \cdot p_f\right)^2} \\
& + \frac{\left(p_\phi \cdot p_i\right) \left(\omega_\phi (E_f - E_i) - 3 m_e^2\right) - m_e^2 \left(p_f \cdot p_i + 2 E_f (\omega_\phi - E_i)\right) + m_e^4}{\left(p_\phi \cdot p_i\right)^2} \\
& + \frac{\left(p_\phi \cdot p_f\right) \left(2 m_e^2 - p_\phi \cdot p_i\right)}{2 \left(p_\phi \cdot p_i\right)^2} \\
&+ \frac{2 \left(\omega_\phi^2 + 2 m_e^2\right) \left(p_f \cdot p_i\right) - \left(p_\phi \cdot p_i\right) \left(p_\phi \cdot p_i + 2 \omega_\phi (E_f - E_i) - 6 m_e^2\right)}{2 \left(p_\phi \cdot p_f\right) \left(p_\phi \cdot p_i\right)} \\
& + \frac{2 m_e^2 \left(\omega_\phi + 2 E_f\right) \left(\omega_\phi - 2 E_i\right) - 4 m_e^4}{2 \left(p_\phi \cdot p_f\right) \left(p_\phi \cdot p_i\right)} + 1\bigg]\,,
\end{aligned}
\end{equation}

\begin{equation}
\mathcal{T}_{\rm ps}={g^{\prime\, 2}_e}\times\frac{
    2 \omega_\phi^2 \left(p_f \cdot p_i - m_e^2\right) + 
    \left(p_\phi \cdot p_f - p_\phi \cdot p_i\right) 
    \left(-p_\phi \cdot p_f + p_\phi \cdot p_i + 2 \omega_\phi (E_f - E_i)\right)
}{
    2 \left(p_\phi \cdot p_f\right) \left(p_\phi \cdot p_i\right)
}\,,
\end{equation}

\begin{equation}
\begin{aligned}
\mathcal{T}_{\rm v}&=\frac{1}{2 \left( \left( p_\phi \cdot p_f \right)^2  \left( p_\phi \cdot p_i \right)^2 \right)} \times \\
&\Bigg( 
    \left( p_\phi \cdot p_f \right) \left( p_\phi \cdot p_i \right) \Big( 
        {I^e}^2 \Big( \left( p_\phi \cdot p_i \right)^2 + 2 \omega_\phi^2 \left( m_e^2 - p_f \cdot p_i \right) 
        + 2 \omega_\phi (E_f - E_i) \left( p_\phi \cdot p_i \right) \Big) \\
    &+ 2 \left( p_\phi \cdot {I^e} \right)^2 \left( -p_f \cdot p_i + 2 E_f E_i + m_e^2 \right) \\
    &+ 2 \left( p_\phi \cdot {I^e} \right) \Big( 
        \left( p_f \cdot p_i \right) \left( -p_f \cdot {I^e} + p_i \cdot {I^e} + 2 \omega_\phi I^e_0 \right) \\
    & + \left( 2 E_f E_i + m_e^2 \right) \left( p_f \cdot {I^e} - p_i \cdot {I^e} \right) \\
    & + \left( p_\phi \cdot p_i \right) \left( p_f \cdot {I^e} - 2 E_f I^e_0 \right) 
        - 2 \omega_\phi m_e^2 I^e_0 \Big) \\
    &+ 2 \left( p_f \cdot {I^e} \right) \Big( 
        2 \left( p_i \cdot {I^e} \right) \left( p_\phi \cdot p_i + p_f \cdot p_i - E_i (\omega_\phi + 2 E_f) 
        + \omega_\phi (\omega_\phi + E_f) - m_e^2 \right) \\
    &+ \left( p_\phi \cdot p_i \right) \left( p_f \cdot {I^e} - 2 I^e_0 (\omega_\phi + E_f + E_i) \right) \Big) 
    \Big) \\
    &- 2 \left( p_\phi \cdot p_f \right)^2 \Big( 
        \left( {I^e}^2 - 2 {I^e_0}^2 \right) \left( p_\phi \cdot p_i \right)^2 \\
    & + \left( p_i \cdot {I^e} \right) \left( p_i \cdot {I^e} - p_\phi \cdot {I^e} \right) 
        \left( p_f \cdot p_i + 2 \omega_\phi E_f - 2 E_f E_i - m_e^2 \right) \\
    & + \left( p_\phi \cdot p_i \right) \Big( 
        \left( p_i \cdot {I^e} \right) \left( 2 \left( p_f \cdot {I^e} \right) + p_i \cdot {I^e} + 2 I^e_0 (\omega_\phi - E_f - E_i) \right) \\
    & + \omega_\phi {I^e}^2 (E_f - E_i) 
        + \left( p_\phi \cdot {I^e} \right) \left( 2 E_i I^e_0 - p_i \cdot {I^e} \right) \Big) \Big) \\
    &- 2 \left( p_f \cdot {I^e} \right) \left( p_\phi \cdot {I^e} + p_f \cdot {I^e} \right) 
        \left( p_\phi \cdot p_i \right)^2 
        \left( p_\phi \cdot p_i + p_f \cdot p_i - 2 E_i (\omega_\phi + E_f) - m_e^2 \right) \\
    &+ \Big( 2 \left( p_i \cdot {I^e} \right) \left( p_i \cdot {I^e} - p_\phi \cdot I^e \right) 
        + {I^e}^2 \left( p_\phi \cdot p_i \right) \Big) 
        \left( p_\phi \cdot p_f \right)^3 
\Bigg)\,,
\end{aligned}
\end{equation}

\begin{equation}
\begin{aligned}
\mathcal{T}_{\rm pv}&=\frac{1}{4 \, (p_\phi \cdot p_f)^2 \, (p_\phi \cdot p_i)^2} \times \\
&\Bigg[  -2 (p_\phi \cdot p_f) (p_\phi \cdot p_i) 
\Bigg[ 2 \Bigg( 2 m_e^2 
\Big[ (p_f \cdot J^e)^2 - 2 J^e_0 (\omega_\phi + E_f - E_i) (p_f \cdot J^e) \\
& + 2 {J^e_0}^2 (m_e^2 - p_f \cdot p_i) \Big] 
+ m_e^2 (p_i \cdot J^e)^2 
- (p_i \cdot J^e) 
\Big[ (p_f \cdot J^e) (p_f \cdot p_i 
- E_i (\omega_\phi + 2 E_f) \\
& + \omega_\phi (\omega_\phi + E_f) + m_e^2) 
- 2 m_e^2 J^e_0 (\omega_\phi + E_f - E_i) \Big] 
+ (p_\phi \cdot J^e)^2 (p_f \cdot p_i - 2 E_f E_i + m_e^2) \\
& + {J^e}^2 
\Big[ (\omega_\phi^2 + 2 m_e^2) (p_f \cdot p_i) 
+ m_e^2 (\omega_\phi + 2 E_f)(\omega_\phi - 2 E_i) - 2 m_e^4 \Big] \\
& + (p_\phi \cdot J^e) 
\Big[ (p_f \cdot p_i) (p_f \cdot J^e 
- p_i \cdot J^e - 2 \omega_\phi J^e_0) \\
& - (2 E_f E_i - 3 m_e^2) (p_f \cdot J^e - p_i \cdot J^e) 
- 2 m_e^2 J^e_0 (\omega_\phi + 2 E_f - 2 E_i) \Big] \Bigg) \\
& - {J^e}^2 (p_\phi \cdot p_i)^2 
+ 2 (p_\phi \cdot p_i) 
\Big[ {J^e}^2 (-\omega_\phi E_f + \omega_\phi E_i + 3 m_e^2) \\
& + (p_f \cdot J^e) 
\big( - (p_f \cdot J^e) 
- 2 (p_i \cdot J^e) 
+ 2 J^e_0 (\omega_\phi + E_f + E_i) \big) \\
& + (p_\phi \cdot J^e) (2 E_f J^e_0 - p_f \cdot J^e) 
- 4 m_e^2 {J^e_0}^2 \Big] \Bigg] \\
& + (p_\phi \cdot p_f)^2 
\Bigg[ (8 {J^e_0}^2 - 4 {J^e}^2) (p_\phi \cdot p_i)^2 
+ 4 (p_\phi \cdot p_i) 
\Big[ {J^e}^2 (-\omega_\phi E_f + \omega_\phi E_i + 3 m_e^2) \\
& - (p_i \cdot J^e) 
\big( 2 (p_f \cdot J^e) + (p_i \cdot J^e) 
+ 2 J^e_0 (\omega_\phi - E_f - E_i) \big) \\
& + (p_\phi \cdot J^e) (p_i \cdot J^e - 2 E_i J^e_0) 
- 4 m_e^2 {J^e_0}^2 \Big] \\
& + 4 (m_e^2 {J^e}^2 
+ (p_i \cdot J^e) (p_\phi \cdot J^e - p_i \cdot J^e)) 
(p_f \cdot p_i + 2 \omega_\phi E_f - 2 E_f E_i - m_e^2) \Bigg] \\
& - 4 (p_\phi \cdot p_i)^2 
\Big[ (p_f \cdot J^e) (p_\phi \cdot J^e + p_f \cdot J^e) 
- m_e^2 {J^e}^2 \Big] \\
& (p_\phi \cdot p_i + p_f \cdot p_i 
- 2 E_i (\omega_\phi + E_f) - m_e^2) \\
& + 2 (p_\phi \cdot p_f)^3 
\Big[ {J^e}^2 (p_\phi \cdot p_i - 2 m_e^2) 
+ 2 (p_i \cdot J^e) 
(p_i \cdot J^e - p_\phi \cdot J^e) \Big] \Bigg]\,,
\end{aligned}
\end{equation}

\begin{equation}
\begin{aligned}
\mathcal{T}_{\rm s-v}&=g_e\,\frac{1}{(p_\phi \cdot p_f)^2 (p_\phi \cdot p_i)^2} 
 \times \\
&m_e\Bigg[  (p_\phi \cdot p_f)^2 
\Big[ (p_\phi \cdot I - 2 (p_i \cdot I)) 
(p_f \cdot p_i + 2 \omega_\phi E_f - 2 E_f E_i - m_e^2) \\
& + (p_\phi \cdot p_i) 
(p_\phi \cdot I - 2 (p_f \cdot I + 2 (p_i \cdot I) 
+ I_0 (\omega_\phi - E_f - E_i))) \Big] \\
& + (p_\phi \cdot I + 2 (p_f \cdot I)) 
(p_\phi \cdot p_i)^2 
\Big[ - (p_\phi \cdot p_i) - (p_f \cdot p_i) 
+ 2 E_i (\omega_\phi + E_f) + m_e^2 \Big] \\
& - (p_\phi \cdot I - 2 (p_i \cdot I)) 
(p_\phi \cdot p_f)^3 \\
& + (p_\phi \cdot p_f) (p_\phi \cdot p_i) 
\Big[ 2 (p_f \cdot I + p_i \cdot I) 
(p_f \cdot p_i - E_i (\omega_\phi + 2 E_f) 
+ \omega_\phi (\omega_\phi + E_f) - m_e^2) \\
& + (p_\phi \cdot p_i) 
(p_\phi \cdot I + 4 (p_f \cdot I) 
+ 2 (p_i \cdot I) - 2 I_0 (\omega_\phi + E_f + E_i)) \Big] 
\Bigg]\,,
\end{aligned}
\end{equation}

\begin{equation}
\begin{aligned}
\mathcal{T}_{\rm ps-pv}&=g^{\prime}_e\,\frac{1}{(p_\phi \cdot p_f)^2 (p_\phi \cdot p_i)^2} \times \\
&m_e \Bigg[  (p_\phi \cdot p_f)^2 \Big[ 
(p_\phi \cdot J^e) (p_\phi \cdot p_i + p_f \cdot p_i 
+ 2 \omega_\phi E_f - 2 E_f E_i - m_e^2) \\
& - 2 (p_\phi \cdot p_i) (p_f \cdot J^e 
- p_i \cdot J^e + J^e_0 (\omega_\phi - E_f + E_i)) \Big] \\
& + (p_\phi \cdot J^e) (p_\phi \cdot p_i)^2 
(p_\phi \cdot p_i + p_f \cdot p_i 
- 2 E_i (\omega_\phi + E_f) - m_e^2)  - (p_\phi \cdot J^e) (p_\phi \cdot p_f)^3 \\
& + (p_\phi \cdot p_f) (p_\phi \cdot p_i) \Big[ 
2 J^e_0 \Big( (\omega_\phi - E_f + E_i) (p_\phi \cdot p_i) 
+ 2 \omega_\phi (p_f \cdot p_i - m_e^2) \Big) \\
& + 2 (p_f \cdot J^e - p_i \cdot J^e) 
(p_\phi \cdot p_i + \omega_\phi (\omega_\phi + E_f - E_i)) \\
& + (p_\phi \cdot J^e) \Big( -p_\phi \cdot p_i 
- 2 (p_f \cdot p_i) + 4 E_f E_i + 2 m_e^2 \Big) 
\Big] \Bigg]\,.
\end{aligned}
\end{equation}

In order to gain the physical intuition about the terms written above, we report below their expressions in the non-relativistic and degenerate limits for electrons. Therefore, we assume $|\textbf{p}_i|\approx|\textbf{p}_f|= k_F$, where $k_F=\beta_F E_F$ is the Fermi momentum, with $E_F$ and $\beta_F$ the Fermi energy and velocity ($E_F\gtrsim m_e\gg T\sim\omega_\phi$) respectively. In these conditions, the momentum transfer $|\textbf{q}|^2\approx |\textbf{p}_{f}-\textbf{p}_{i}|^2=2 k_F^2 (1-x_{if})$, with $x_{if}$ the cosine of the angle between $\textbf{p}_{i}$ and $\textbf{p}_{f}$. Therefore, with these approximations, we find
\begin{equation}
|\overline{\mathcal{M}}|^2=\frac{n_{\rm He}\,Z_{\rm He}^2\,16\pi^2 \alpha^2}{4k_F^4(1-\beta_F^2x_{if})(1-\beta_F^2 x_{if}+k_s^2/2k_F^2)}\left(\mathcal{T}_{\rm s} + \mathcal{T}_{\rm ps} + \mathcal{T}_{\rm v} + \mathcal{T}_{\rm pv} + \mathcal{T}_{\rm s-v} + \mathcal{T}_{\rm ps-pv}\right)\,,
\label{eq:MbarNR}
\end{equation}
and
\begin{equation}
\begin{aligned} 
\mathcal{T}_s&=g_e^2\times\frac{\beta_F^2E_F^2}{\omega_\phi^2}\frac{(1-\beta_F^2)(x_{f\phi}-x_{i\phi})^2[2-\beta_F^2(1-x_{if})]}{(1-\beta_F\,x_{f\phi})^2(1-\beta_F x_{i\phi})^2}+\mathcal{O}(E_F/\omega_\phi)\equiv g_e^2\times\frac{\beta_F^2E_F^2(1-\beta_F^2)}{\omega_\phi^2} \mathcal{A}_s\,,
\end{aligned}
\label{eq:sNR}
\end{equation}
\begin{equation}
\mathcal{T}_{\rm ps}={g^{\prime\, 2}_e}\times\beta_F^2\frac{2 (1-x_{if})-(x_{f\phi}-x_{i\phi})^2}{2(1-\beta_F x_{i\phi})(1-\beta_F x_{f\phi})}\equiv{g^{\prime\, 2}_e}\times\frac{\beta_F^2}{2} \mathcal{A}_{\rm ps}\,,
\label{eq:psNR}
\end{equation}
\begin{equation}
\begin{aligned}
\mathcal{T}_{\rm v} & =|\textbf{I}^e|^2 \frac{\beta_F^2\,E_F^2 \left[2-\beta_F^2 (1-x_{if})\right] \left[(1-\beta_F x_{f\phi})x_{iI}-(1-\beta_F x_{i\phi})x_{fI}\right]^2}{\omega_\phi^2 (1-\beta_F x_{i\phi})^2(1-\beta_F x_{f\phi})^2} \\
&+ 2|\textbf{I}^e| I_0^e \frac{\beta_F^2\, E_F^2 (x_{i\phi}-x_{f\phi}) \left[2-\beta_F^2 (1-x_{if})\right] \left[(1-\beta_F x_{fI} x_{i\phi})x_{fI}-(1-\beta_F x_{f\phi})x_{iI} \right]}{\omega_\phi^2 (1-\beta_F x_{i\phi})^2(1-\beta_F x_{f\phi})^2} \\
&+{I_0^e}^2 \frac{\beta_F^2\, E_F^2 (x_{f\phi}-x_{i\phi})^2\left[2-\beta_F^2(1-x_{if})\right]}{\omega_\phi^2 (1-\beta_F x_{i\phi})^2(1-\beta_F x_{f\phi})^2} + \mathcal{O}(E_F/\omega_\phi) \equiv \frac{\beta_F^2 E_F^2}{\omega_\phi^2} \mathcal{A}_{\rm v} (|\textbf{I}^e|,\, I_0^e)\,,
\end{aligned}
\label{eq:vNR}
\end{equation}
\begin{equation}
\begin{aligned}
\mathcal{T}_{\rm pv}&= |\textbf{J}^e|^2 \frac{\beta_F^2 E_F^2}{\omega_\phi^2 (1-\beta_F x_{i\phi})^2(1-\beta_F x_{f\phi})^2} \bigg[x_{f\phi}^2 \left(2-\beta_F^2 (1-x_{if})\right) \left(1-\beta_F^2 \left(1-x_{iJ}^2\right)\right)\\
&+2 x_{f\phi} \bigg(x_{i\phi} \beta_F^4 \left(x_{fJ}^2-x_{fJ} (x_{if}+1) x_{iJ}+x_{if}+x_{iJ}^2-1\right)\\
&-x_{i\phi}\left(\beta_F^2 \left(x_{fJ}^2+x_{if}+x_{iJ}^2-3\right)-2\right) \\
&-\beta_F (x_{fJ}-x_{iJ}) \left(\beta_F^2 (x_{fJ}-x_{if} x_{iJ})-x_{fJ}-x_{iJ}\right)\bigg)\\
&+x_{i\phi}^2 \left(\beta_F^2 \left(x_{fJ}^2-1\right)+1\right) \left(\beta_F^2 (x_{if}-1)+2\right) \\
&-2 \beta_F x_{i\phi} (x_{fJ}-x_{iJ}) \left(\beta_F^2 (x_{fJ} x_{if}-x_{iJ})+x_{fJ}+x_{iJ}\right)+\beta_F^2 (x_{if}+1) (x_{fJ}-x_{iJ})^2\bigg] \\
& + 2 J_0^e |\textbf{J}^e| \frac{\beta_F^2 E_F^2 (x_{f\phi}-x_{i\phi}) \left(2-\beta_F^2 (1-x_{if})\right) \left[(1-\beta_F\,x_{i\phi})x_{fJ}-(1-\beta_F x_{f\phi})x_{iJ}\right]}{\omega_\phi^2 (1-\beta_F x_{i\phi})^2(1-\beta_F x_{f\phi})^2} \\ 
&+{J_0^e}^2 \frac{\beta_F^2 E_F^2}{\omega_\phi^2 (1-\beta_F x_{i\phi})^2(1-\beta_F x_{f\phi})^2} \bigg[2 \beta_F^2 \left(x_{f\phi}^2-2 x_{f\phi} x_{i\phi} x_{if}+x_{i\phi}^2+2 x_{if}-2\right) \\
&+\beta_F^4 (x_{if}-1) (x_{f\phi}+x_{i\phi})^2-4 \beta_F^3 (x_{if}-1) (x_{f\phi}+x_{i\phi}) \\
&+4 \beta_F (x_{if}-1) (x_{f\phi}+x_{i\phi})+4(1- x_{if})\bigg] + \mathcal{O}(E_F/\omega_\phi) \equiv \frac{\beta_F^2 E_F^2}{\omega_\phi^2} \mathcal{A}_{\rm pv} (|\textbf{J}^e|,\, J_0^e)\,,
\end{aligned}
\label{eq:pvNR}
\end{equation}
\begin{equation}
\begin{aligned}
\mathcal{T}_{\rm s-v}&=g_e\times  \frac{\beta_F^2 E_F^2}{\omega_\phi^2}\bigg[ \frac{2\,|\textbf{I}^e|  \left(1-\beta_F^2\right) (x_{i\phi}-x_{f\phi}) \left[2-\beta_F^2(1-x_{if})\right] \left[(1-\beta_F x_{i\phi})x_{fI}-(1-\beta_F x_{f\phi})x_{iI}\right]}{(1-\beta_F x_{i\phi})^2(1-\beta_F x_{f\phi})^2} \\ 
& +\frac{2 I_0^e \left(1-\beta_F^2\right) (x_{i\phi}-x_{f\phi})^2 \left[2-\beta_F^2(1-x_{if})\right]}{(1-\beta_F x_{i\phi})^2(1-\beta_F x_{f\phi})^2}\bigg] + \mathcal{O}(E_F/\omega_\phi)\, \equiv g_e\times  \frac{\beta_F^2 E_F^2}{\omega_\phi^2} \mathcal{A}_{\rm s-v}(|\textbf{I}^e|,\, I_0^e),
\end{aligned}
\label{eq:svNR}
\end{equation}
\begin{equation}
\begin{aligned}
\mathcal{T}_{\rm ps-pv}=&g^{\prime}_e\,\times  \frac{\beta_F^2E_F(1-\beta_F^2)}{\omega_\phi (1-\beta_F x_{i\phi})^2(1-\beta_F x_{f\phi})^2}\times\\ 
&\bigg[  |\textbf{J}^e| (x_{i\phi}-x_{f\phi}) \bigg[2 (1-\beta_F x_{f\phi})(1-\beta_F x_{i\phi}) (x_{fJ}-x_{iJ}) \\
&-x_{\phi J} (x_{f\phi}-x_{i\phi}) \left[2-\beta_F^2 (1-x_{if})\right]\bigg] \\
& + J_0^e  \left[(1-x_{if}) [\beta_F^2(x_{f\phi}+x_{i\phi})^2+4 \beta_F (x_{f\phi}+x_{i\phi})-4]+2 (x_{f\phi}-x_{i\phi})^2\right]\bigg] + \mathcal{O}(E_F/\omega_\phi) \\
\equiv & g^{\prime}_e \times  \frac{\beta_F^2E_F(1-\beta_F^2)}{\omega_\phi} \mathcal{A}_{\rm ps-pv}(|\textbf{J}^e|,\, J_0^e)  \,,
\end{aligned}
\label{eq:pspvNR}
\end{equation}
where $x_{ij}$ and $x_{iI(J)}$ are the cosines of the angles between the particles $i$-$j$ and between the particle $i$ and the vector $\textbf{I}^e$($\textbf{J}^e$), i.e. $\textbf{p}_i \cdot \textbf{p}_j = |\textbf{p}_i||\textbf{p}_j| x_{ij}$ and $\textbf{p}_i \cdot \textbf{I}^e = |\textbf{p}_i||\textbf{I}^e| x_{iI}$ respectively. The equations above show that at the leading order the matrix elements for all the interaction but the pseudoscalar and the pseudoscalar-pseudovector terms scale as $E_F^2/\omega_\phi^2$. On the other hand, the pseudoscalar term scales as $(E_F/\omega_\phi)^0$ [see Eq.~\eqref{eq:psNR}] and the interference term $\mathcal{T}_{\rm ps-pv} \propto E_F/\omega_\phi$ [see Eq.~\eqref{eq:pspvNR}]. As further discussed in Appendix~\ref{app:Qdeg}, this explains the reason why the pseudoscalar and the pseudoscalar-pseudovector emissivities are suppressed by a further factor $T^2/m_e^2$ and $T/m_e$ with respect to the other terms. Additionally, these expressions reveal also the reason why there are no terms proportional to $|\textbf{I}^e|$ or $|\textbf{J}^e|$ in the complete emissivity in Eq.~\eqref{eq:epsgen}. Indeed, when we compute the emissivity, we average over the angles between each particle and the vector $\textbf{I}^e$ ($\textbf{J}^e$), implying that all the terms proportional to $x_{iI(J)}$, $x_{fI(J)}$ and $x_{\phi I(J)}$ vanish.

\subsection{Emissivity in the degenerate limit}
\label{app:Qdeg}
In full generality, one can write the emissivity per unit volume as
\begin{equation}
    Q_\phi = \rho\, \varepsilon_\phi=\int\frac{2d^{3}\textbf{p}_{i}}{(2\pi)^{3}2E_{i}}\frac{2d^{3}\textbf{p}_{f}}{(2\pi)^{3}2E_{f}}\frac{d^3 \textbf{p}_\phi}{(2\pi)^{3}2\omega_\phi}\omega_\phi\\
    (2\pi)\delta(E_{i}-E_{f}-\omega_{\phi})\,|\overline{\mathcal{M}}|^{2}f_{i}(1-f_{f})\,,
\label{eq:Qphi}
\end{equation}
with $|\overline{\mathcal{M}}|^2$ given by Eq.~\eqref{eq:Mbar} and $f_{i,f}$ the electron Fermi-Dirac distribution functions at energies $E_i$ and $E_f$. In the limit of massless boson and by integrating over $\omega_\phi$ to remove the $\delta$ function, the expression above can be simplified to
\begin{equation}
     Q_\phi = \frac{1}{2\pi^4}\int dE_i\,dE_f\,\frac{d\Omega_i}{4\pi} \,\frac{d\Omega_f}{4\pi} \frac{d\Omega_\phi}{4\pi} f_{i}(1-f_{f})|\textbf{p}_{i}||\textbf{p}_{f}|\omega_\phi^2|\overline{\mathcal{M}}|^{2}\,,
\label{eq:Qphired}   
\end{equation}
with $\omega_\phi = E_i-E_f$. In the limit of highly-degenerate electrons, the integrals over the electron energies can be done analytically. Moreover, the electron momenta are close to the Fermi surface $|\textbf{p}_{i}|\approx |\textbf{p}_{f}|=k_F$, implying that $|\overline{\mathcal{M}}|^2$ is given by Eq.~\eqref{eq:MbarNR}. Therefore, we can write
\begin{equation}
\begin{split}
Q_\phi = & \frac{2\alpha^2\,n_{\rm He}\,Z_{\rm He}^2}{\pi^3}\int_{m_e}^\infty dE_i\,\int_{m_e}^{E_i}dE_f\,f_{i}(1-f_{f})\omega_\phi^2 \int\frac{d\Omega_i}{4\pi} \,\frac{d\Omega_f}{4\pi} \frac{d\Omega_\phi}{4\pi} \\
& \frac{1}{k_F^2(1-\beta_F^2x_{if})(1-\beta_F^2 x_{if}+k_s^2/2k_F^2)}\left(\mathcal{T}_{\rm s} + \mathcal{T}_{\rm ps} + \mathcal{T}_{\rm v} + \mathcal{T}_{\rm pv} + \mathcal{T}_{\rm s-v} + \mathcal{T}_{\rm ps-pv}\right)\,.
\end{split}
\end{equation}
In these conditions, from Eqs.~\eqref{eq:sNR}-\eqref{eq:pspvNR} it is apparent that $\mathcal{T}_{\rm ps}\propto\omega_\phi^{0}$, $\mathcal{T}_{\rm ps-pv}\propto\omega_\phi^{-1}$ and for all the other cases $\mathcal{T}\propto\omega_\phi^{-2}$. Therefore, from the integrals $\int dE_i dE_f f_1 (1-f_f) (E_i-E_f)^2 \approx \pi^4 T^4/15\,$ for the pseudoscalar case, $\int dE_i dE_f f_1 (1-f_f) (E_i-E_f) \approx 12\, T^3/5$ for the interference term and $\int dE_i dE_f f_1 (1-f_f) \approx \pi^2\,T^2/6$ for all the other cases, we have
\begin{equation}
    Q_{\phi, {\rm ps}} \approx \frac{\pi \alpha^2\,{g_e^\prime}^2\,n_{\rm He}\,Z_{\rm He}^2}{15} \frac{T^4}{m_e^2} \, G_{\rm ps}\,,
\end{equation}
\begin{equation}
    Q_{\phi, {\rm ps-pv}} \approx \frac{24\alpha^2\,g_e^\prime n_{\rm He}\,Z_{\rm He}^2}{5\pi^3} \frac{T^3}{m_e} \, G_{\rm ps-pv}\,,
\end{equation}
\begin{equation}
    Q_{\phi, {\rm i}} \approx \frac{\alpha^2\,n_{\rm He}\,Z_{\rm He}^2}{3 \pi} T^2 \, G_{\rm i}\,,\quad {\rm i = s,\,v,\,pv,\,s-v}
\end{equation}
where $G_{\rm k}$ (${\rm k=s,\, ps,\, v,\, pv,\, s-v,\, ps-pv}$) are non-trivial functions involving integrations over $d\Omega_i, d\Omega_f,\,d\Omega_\phi$:
\begin{equation}
    G_{\rm ps}=\int\frac{d\Omega_i}{4\pi} \,\frac{d\Omega_f}{4\pi} \frac{d\Omega_\phi}{4\pi} \frac{(1-\beta_F^2)\mathcal{A}_{\rm ps}}{(1-\beta_F^2x_{if})(1-\beta_F^2 x_{if}+k_s^2/2k_F^2)},
\end{equation}
\begin{equation}
    G_{\rm ps-pv}=\int\frac{d\Omega_i}{4\pi} \,\frac{d\Omega_f}{4\pi} \frac{d\Omega_\phi}{4\pi} \frac{(1-\beta_F^2)\mathcal{A}_{\rm ps-pv}}{(1-\beta_F^2x_{if})(1-\beta_F^2 x_{if}+k_s^2/2k_F^2)},
\end{equation}
\begin{equation}
    G_{\rm s}=\int\frac{d\Omega_i}{4\pi} \,\frac{d\Omega_f}{4\pi} \frac{d\Omega_\phi}{4\pi} \frac{g_e^2\,(1-\beta_F^2)\,\mathcal{A}_{\rm s}}{(1-\beta_F^2x_{if})(1-\beta_F^2 x_{if}+k_s^2/2k_F^2)}\,,
\end{equation}
\begin{equation}
    G_{\rm i}=\int\frac{d\Omega_i}{4\pi} \,\frac{d\Omega_f}{4\pi} \frac{d\Omega_\phi}{4\pi} \frac{\mathcal{A}_{\rm i}}{(1-\beta_F^2x_{if})(1-\beta_F^2 x_{if}+k_s^2/2k_F^2)}\,,\quad {\rm i = v,\,pv,\,s-v}\,.
\end{equation}
We stress that with these aproximations the expression of the emissivities agree with literature for the scalar~\cite{Bottaro:2023gep} and pseudoscalar~\cite{Raffelt:1989zt} cases.

\bibliographystyle{utphys}
\bibliography{references}

\providecommand{\href}[2]{#2}\begingroup\raggedright\begin{thebibliography}{100}

\bibitem{Wu:1957my}
C.~S. Wu, E.~Ambler, R.~W. Hayward, D.~D. Hoppes, and R.~P. Hudson,
  ``{Experimental Test of Parity Conservation in $\beta$ Decay},''
  \href{http://dx.doi.org/10.1103/PhysRev.105.1413}{{\em Phys. Rev.} {\bfseries
  105} (1957) 1413--1414}.

\bibitem{GargamelleNeutrino:1973jyy}
{\bfseries Gargamelle Neutrino} Collaboration, F.~J. Hasert {\em et~al.},
  ``{Observation of Neutrino Like Interactions Without Muon Or Electron in the
  Gargamelle Neutrino Experiment},''
  \href{http://dx.doi.org/10.1016/0370-2693(73)90499-1}{{\em Phys. Lett. B}
  {\bfseries 46} (1973) 138--140}.

\bibitem{UA1:1983crd}
{\bfseries UA1} Collaboration, G.~Arnison {\em et~al.}, ``{Experimental
  Observation of Isolated Large Transverse Energy Electrons with Associated
  Missing Energy at $\sqrt{s}= 540$ GeV},''
  \href{http://dx.doi.org/10.1016/0370-2693(83)91177-2}{{\em Phys. Lett. B}
  {\bfseries 122} (1983) 103--116}.

\bibitem{UA1:1983mne}
{\bfseries UA1} Collaboration, G.~Arnison {\em et~al.}, ``{Experimental
  Observation of Lepton Pairs of Invariant Mass Around 95-GeV/c**2 at the CERN
  SPS Collider},'' \href{http://dx.doi.org/10.1016/0370-2693(83)90188-0}{{\em
  Phys. Lett. B} {\bfseries 126} (1983) 398--410}.

\bibitem{Kostelecky:2008ts}
V.~A. Kostelecky and N.~Russell, ``{Data Tables for Lorentz and CPT
  Violation},'' \href{http://dx.doi.org/10.1103/RevModPhys.83.11}{{\em Rev.
  Mod. Phys.} {\bfseries 83} (2011) 11--31},
  \href{http://arxiv.org/abs/0801.0287}{{\ttfamily arXiv:0801.0287 [hep-ph]}}.

\bibitem{Gupta:2022qoq}
R.~S. Gupta, J.~Jaeckel, and M.~Spannowsky, ``{Probing Poincar\'e violation},''
  \href{http://dx.doi.org/10.1007/JHEP11(2023)026}{{\em JHEP} {\bfseries 11}
  (2023) 026}, \href{http://arxiv.org/abs/2211.04490}{{\ttfamily
  arXiv:2211.04490 [hep-ph]}}.

\bibitem{Colladay:1996iz}
D.~Colladay and V.~A. Kostelecky, ``{CPT violation and the standard model},''
  \href{http://dx.doi.org/10.1103/PhysRevD.55.6760}{{\em Phys. Rev. D}
  {\bfseries 55} (1997) 6760--6774},
  \href{http://arxiv.org/abs/hep-ph/9703464}{{\ttfamily arXiv:hep-ph/9703464}}.

\bibitem{Colladay:1998fq}
D.~Colladay and V.~A. Kostelecky, ``{Lorentz violating extension of the
  standard model},'' \href{http://dx.doi.org/10.1103/PhysRevD.58.116002}{{\em
  Phys. Rev. D} {\bfseries 58} (1998) 116002},
  \href{http://arxiv.org/abs/hep-ph/9809521}{{\ttfamily arXiv:hep-ph/9809521}}.

\bibitem{Kostelecky:2003fs}
V.~A. Kostelecky, ``{Gravity, Lorentz violation, and the standard model},''
  \href{http://dx.doi.org/10.1103/PhysRevD.69.105009}{{\em Phys. Rev. D}
  {\bfseries 69} (2004) 105009},
  \href{http://arxiv.org/abs/hep-th/0312310}{{\ttfamily arXiv:hep-th/0312310}}.

\bibitem{Kostelecky:2000mm}
V.~A. Kostelecky and R.~Lehnert, ``{Stability, causality, and Lorentz and CPT
  violation},'' \href{http://dx.doi.org/10.1103/PhysRevD.63.065008}{{\em Phys.
  Rev. D} {\bfseries 63} (2001) 065008},
  \href{http://arxiv.org/abs/hep-th/0012060}{{\ttfamily arXiv:hep-th/0012060}}.

\bibitem{Kostelecky:2024rsn}
A.~Kostelecky, R.~Lehnert, M.~Schreck, and B.~Seradjeh, ``{Physical
  interpretation of large Lorentz violation via Weyl semimetals},''
  \href{http://arxiv.org/abs/2412.18034}{{\ttfamily arXiv:2412.18034
  [hep-ph]}}.

\bibitem{Kostelecky:2001jc}
V.~A. Kostelecky, C.~D. Lane, and A.~G.~M. Pickering, ``{One loop
  renormalization of Lorentz violating electrodynamics},''
  \href{http://dx.doi.org/10.1103/PhysRevD.65.056006}{{\em Phys. Rev. D}
  {\bfseries 65} (2002) 056006},
  \href{http://arxiv.org/abs/hep-th/0111123}{{\ttfamily arXiv:hep-th/0111123}}.

\bibitem{Colladay:2007aj}
D.~Colladay and P.~McDonald, ``{One-Loop Renormalization of QCD with Lorentz
  Violation},'' \href{http://dx.doi.org/10.1103/PhysRevD.77.085006}{{\em Phys.
  Rev. D} {\bfseries 77} (2008) 085006},
  \href{http://arxiv.org/abs/0712.2055}{{\ttfamily arXiv:0712.2055 [hep-ph]}}.

\bibitem{Colladay:2009rb}
D.~Colladay and P.~McDonald, ``{One-Loop Renormalization of the Electroweak
  Sector with Lorentz Violation},''
  \href{http://dx.doi.org/10.1103/PhysRevD.79.125019}{{\em Phys. Rev. D}
  {\bfseries 79} (2009) 125019},
  \href{http://arxiv.org/abs/0904.1219}{{\ttfamily arXiv:0904.1219 [hep-ph]}}.

\bibitem{Kostelecky:1988zi}
V.~A. Kostelecky and S.~Samuel, ``{Spontaneous Breaking of Lorentz Symmetry in
  String Theory},'' \href{http://dx.doi.org/10.1103/PhysRevD.39.683}{{\em Phys.
  Rev. D} {\bfseries 39} (1989) 683}.

\bibitem{Kostelecky:1991ak}
V.~A. Kostelecky and R.~Potting, ``{CPT and strings},''
  \href{http://dx.doi.org/10.1016/0550-3213(91)90071-5}{{\em Nucl. Phys. B}
  {\bfseries 359} (1991) 545--570}.

\bibitem{Altschul:2005mu}
B.~Altschul and V.~A. Kostelecky, ``{Spontaneous Lorentz violation and
  nonpolynomial interactions},''
  \href{http://dx.doi.org/10.1016/j.physletb.2005.09.018}{{\em Phys. Lett. B}
  {\bfseries 628} (2005) 106--112},
  \href{http://arxiv.org/abs/hep-th/0509068}{{\ttfamily arXiv:hep-th/0509068}}.

\bibitem{Gambini:1998it}
R.~Gambini and J.~Pullin, ``{Nonstandard optics from quantum space-time},''
  \href{http://dx.doi.org/10.1103/PhysRevD.59.124021}{{\em Phys. Rev. D}
  {\bfseries 59} (1999) 124021},
  \href{http://arxiv.org/abs/gr-qc/9809038}{{\ttfamily arXiv:gr-qc/9809038}}.

\bibitem{Alfaro:2001rb}
J.~Alfaro, H.~A. Morales-Tecotl, and L.~F. Urrutia, ``{Loop quantum gravity and
  light propagation},''
  \href{http://dx.doi.org/10.1103/PhysRevD.65.103509}{{\em Phys. Rev. D}
  {\bfseries 65} (2002) 103509},
  \href{http://arxiv.org/abs/hep-th/0108061}{{\ttfamily arXiv:hep-th/0108061}}.

\bibitem{Kostelecky:2002ca}
V.~A. Kostelecky, R.~Lehnert, and M.~J. Perry, ``{Spacetime - varying couplings
  and Lorentz violation},''
  \href{http://dx.doi.org/10.1103/PhysRevD.68.123511}{{\em Phys. Rev. D}
  {\bfseries 68} (2003) 123511},
  \href{http://arxiv.org/abs/astro-ph/0212003}{{\ttfamily
  arXiv:astro-ph/0212003}}.

\bibitem{Ferrero:2009jb}
A.~Ferrero and B.~Altschul, ``{Radiatively Induced Lorentz and Gauge Symmetry
  Violation in Electrodynamics with Varying alpha},''
  \href{http://dx.doi.org/10.1103/PhysRevD.80.125010}{{\em Phys. Rev. D}
  {\bfseries 80} (2009) 125010},
  \href{http://arxiv.org/abs/0910.5202}{{\ttfamily arXiv:0910.5202 [hep-th]}}.

\bibitem{Mocioiu:2000ip}
I.~Mocioiu, M.~Pospelov, and R.~Roiban, ``{Low-energy limits on the
  antisymmetric tensor field background on the brane and on the noncommutative
  scale},'' \href{http://dx.doi.org/10.1016/S0370-2693(00)00928-X}{{\em Phys.
  Lett. B} {\bfseries 489} (2000) 390--396},
  \href{http://arxiv.org/abs/hep-ph/0005191}{{\ttfamily arXiv:hep-ph/0005191}}.

\bibitem{Carroll:2001ws}
S.~M. Carroll, J.~A. Harvey, V.~A. Kostelecky, C.~D. Lane, and T.~Okamoto,
  ``{Noncommutative field theory and Lorentz violation},''
  \href{http://dx.doi.org/10.1103/PhysRevLett.87.141601}{{\em Phys. Rev. Lett.}
  {\bfseries 87} (2001) 141601},
  \href{http://arxiv.org/abs/hep-th/0105082}{{\ttfamily arXiv:hep-th/0105082}}.

\bibitem{Altschul:2006jj}
B.~Altschul, ``{Lorentz Violation and the Yukawa Potential},''
  \href{http://dx.doi.org/10.1016/j.physletb.2006.07.021}{{\em Phys. Lett. B}
  {\bfseries 639} (2006) 679--683},
  \href{http://arxiv.org/abs/hep-th/0605044}{{\ttfamily arXiv:hep-th/0605044}}.

\bibitem{Ferrero:2011yu}
A.~Ferrero and B.~Altschul, ``{Renormalization of Scalar and Yukawa Field
  Theories with Lorentz Violation},''
  \href{http://dx.doi.org/10.1103/PhysRevD.84.065030}{{\em Phys. Rev. D}
  {\bfseries 84} (2011) 065030},
  \href{http://arxiv.org/abs/1104.4778}{{\ttfamily arXiv:1104.4778 [hep-th]}}.

\bibitem{Altschul:2012xu}
B.~Altschul, ``{Lorentz and CPT Violation in Scalar-Mediated Potentials},''
  \href{http://dx.doi.org/10.1103/PhysRevD.87.045012}{{\em Phys. Rev. D}
  {\bfseries 87} no.~4, (2013) 045012},
  \href{http://arxiv.org/abs/1211.6614}{{\ttfamily arXiv:1211.6614 [hep-th]}}.

\bibitem{Altschul:2014gqa}
B.~Altschul, ``{Lorentz Violation in Fermion-Antifermion Decays of Spinless
  Particles},'' \href{http://dx.doi.org/10.1103/PhysRevD.89.116007}{{\em Phys.
  Rev. D} {\bfseries 89} no.~11, (2014) 116007},
  \href{http://arxiv.org/abs/1403.2751}{{\ttfamily arXiv:1403.2751 [hep-ph]}}.

\bibitem{Heckel:2006ww}
B.~R. Heckel, C.~E. Cramer, T.~S. Cook, E.~G. Adelberger, S.~Schlamminger, and
  U.~Schmidt, ``{New CP-violation and preferred-frame tests with polarized
  electrons},'' \href{http://dx.doi.org/10.1103/PhysRevLett.97.021603}{{\em
  Phys. Rev. Lett.} {\bfseries 97} (2006) 021603},
  \href{http://arxiv.org/abs/hep-ph/0606218}{{\ttfamily arXiv:hep-ph/0606218}}.

\bibitem{Heckel:2008hw}
B.~R. Heckel, E.~G. Adelberger, C.~E. Cramer, T.~S. Cook, S.~Schlamminger, and
  U.~Schmidt, ``{Preferred-Frame and CP-Violation Tests with Polarized
  Electrons},'' \href{http://dx.doi.org/10.1103/PhysRevD.78.092006}{{\em Phys.
  Rev. D} {\bfseries 78} (2008) 092006},
  \href{http://arxiv.org/abs/0808.2673}{{\ttfamily arXiv:0808.2673 [hep-ex]}}.

\bibitem{Bluhm:1999ev}
R.~Bluhm and V.~A. Kostelecky, ``{Lorentz and CPT tests with spin polarized
  solids},'' \href{http://dx.doi.org/10.1103/PhysRevLett.84.1381}{{\em Phys.
  Rev. Lett.} {\bfseries 84} (2000) 1381--1384},
  \href{http://arxiv.org/abs/hep-ph/9912542}{{\ttfamily arXiv:hep-ph/9912542}}.

\bibitem{Hees:2018fpg}
A.~Hees, O.~Minazzoli, E.~Savalle, Y.~V. Stadnik, and P.~Wolf, ``{Violation of
  the equivalence principle from light scalar dark matter},''
  \href{http://dx.doi.org/10.1103/PhysRevD.98.064051}{{\em Phys. Rev. D}
  {\bfseries 98} no.~6, (2018) 064051},
  \href{http://arxiv.org/abs/1807.04512}{{\ttfamily arXiv:1807.04512 [gr-qc]}}.

\bibitem{Adelberger:2003zx}
E.~G. Adelberger, B.~R. Heckel, and A.~E. Nelson, ``{Tests of the gravitational
  inverse square law},''
  \href{http://dx.doi.org/10.1146/annurev.nucl.53.041002.110503}{{\em Ann. Rev.
  Nucl. Part. Sci.} {\bfseries 53} (2003) 77--121},
  \href{http://arxiv.org/abs/hep-ph/0307284}{{\ttfamily arXiv:hep-ph/0307284}}.

\bibitem{Fischbach:1996eq}
E.~Fischbach and C.~Talmadge, ``{Ten years of the fifth force},'' in {\em {31st
  Rencontres de Moriond: Dark Matter and Cosmology, Quantum Measurements and
  Experimental Gravitation}}, pp.~443--451.
\newblock 1996.
\newblock \href{http://arxiv.org/abs/hep-ph/9606249}{{\ttfamily
  arXiv:hep-ph/9606249}}.

\bibitem{KONOPLIV2011401}
A.~S. Konopliv, S.~W. Asmar, W.~M. Folkner, Özgür Karatekin, D.~C. Nunes,
  S.~E. Smrekar, C.~F. Yoder, and M.~T. Zuber, ``Mars high resolution gravity
  fields from mro, mars seasonal gravity, and other dynamical parameters,''
  \href{http://dx.doi.org/https://doi.org/10.1016/j.icarus.2010.10.004}{{\em
  Icarus} {\bfseries 211} no.~1, (2011) 401--428}.
  \url{https://www.sciencedirect.com/science/article/pii/S0019103510003830}.

\bibitem{Berge:2017ovy}
J.~Berg\'e, P.~Brax, G.~M\'etris, M.~Pernot-Borr\`as, P.~Touboul, and J.-P.
  Uzan, ``{MICROSCOPE Mission: First Constraints on the Violation of the Weak
  Equivalence Principle by a Light Scalar Dilaton},''
  \href{http://dx.doi.org/10.1103/PhysRevLett.120.141101}{{\em Phys. Rev.
  Lett.} {\bfseries 120} no.~14, (2018) 141101},
  \href{http://arxiv.org/abs/1712.00483}{{\ttfamily arXiv:1712.00483 [gr-qc]}}.

\bibitem{Brzeminski:2022sde}
D.~Brzeminski, Z.~Chacko, A.~Dev, I.~Flood, and A.~Hook, ``{Searching for a
  fifth force with atomic and nuclear clocks},''
  \href{http://dx.doi.org/10.1103/PhysRevD.106.095031}{{\em Phys. Rev. D}
  {\bfseries 106} no.~9, (2022) 095031},
  \href{http://arxiv.org/abs/2207.14310}{{\ttfamily arXiv:2207.14310
  [hep-ph]}}.

\bibitem{PhysRevLett.89.253002}
T.~W. Kornack and M.~V. Romalis, ``Dynamics of two overlapping spin ensembles
  interacting by spin exchange,''
  \href{http://dx.doi.org/10.1103/PhysRevLett.89.253002}{{\em Phys. Rev. Lett.}
  {\bfseries 89} (Dec, 2002) 253002}.
  \url{https://link.aps.org/doi/10.1103/PhysRevLett.89.253002}.

\bibitem{Kornack:2005xrn}
T.~W. Kornack, R.~K. Ghosh, and M.~V. Romalis, ``{Nuclear Spin Gyroscope Based
  on an Atomic Comagnetometer},''
  \href{http://dx.doi.org/10.1103/PhysRevLett.95.230801}{{\em Phys. Rev. Lett.}
  {\bfseries 95} no.~23, (2005) 230801}.

\bibitem{Brown:2010dt}
J.~M. Brown, S.~J. Smullin, T.~W. Kornack, and M.~V. Romalis, ``{New limit on
  Lorentz and CPT-violating neutron spin interactions},''
  \href{http://dx.doi.org/10.1103/PhysRevLett.105.151604}{{\em Phys. Rev.
  Lett.} {\bfseries 105} (2010) 151604},
  \href{http://arxiv.org/abs/1006.5425}{{\ttfamily arXiv:1006.5425
  [physics.atom-ph]}}.

\bibitem{Hoedl:2011zz}
S.~A. Hoedl, F.~Fleischer, E.~G. Adelberger, and B.~R. Heckel, ``{Improved
  Constraints on an Axion-Mediated Force},''
  \href{http://dx.doi.org/10.1103/PhysRevLett.106.041801}{{\em Phys. Rev.
  Lett.} {\bfseries 106} (2011) 041801}.

\bibitem{Peck:2012pt}
S.~K. Peck, D.~K. Kim, D.~Stein, D.~Orbaker, A.~Foss, M.~T. Hummon, and L.~R.
  Hunter, ``{Limits on local Lorentz invariance in mercury and cesium},''
  \href{http://dx.doi.org/10.1103/PhysRevA.86.012109}{{\em Phys. Rev. A}
  {\bfseries 86} (2012) 012109},
  \href{http://arxiv.org/abs/1205.5022}{{\ttfamily arXiv:1205.5022
  [physics.atom-ph]}}.

\bibitem{Allmendinger:2013eya}
F.~Allmendinger, W.~Heil, S.~Karpuk, W.~Kilian, A.~Scharth, U.~Schmidt,
  A.~Schnabel, Y.~Sobolev, and K.~Tullney, ``{New Limit on Lorentz-Invariance-
  and CPT-Violating Neutron Spin Interactions Using a Free-Spin-Precession
  $^3$He - $^{129}$Xe Comagnetometer},''
  \href{http://dx.doi.org/10.1103/PhysRevLett.112.110801}{{\em Phys. Rev.
  Lett.} {\bfseries 112} no.~11, (2014) 110801},
  \href{http://arxiv.org/abs/1312.3225}{{\ttfamily arXiv:1312.3225 [gr-qc]}}.

\bibitem{Budker:2013hfa}
D.~Budker, P.~W. Graham, M.~Ledbetter, S.~Rajendran, and A.~Sushkov,
  ``{Proposal for a Cosmic Axion Spin Precession Experiment (CASPEr)},''
  \href{http://dx.doi.org/10.1103/PhysRevX.4.021030}{{\em Phys. Rev. X}
  {\bfseries 4} no.~2, (2014) 021030},
  \href{http://arxiv.org/abs/1306.6089}{{\ttfamily arXiv:1306.6089 [hep-ph]}}.

\bibitem{Kahn:2016aff}
Y.~Kahn, B.~R. Safdi, and J.~Thaler, ``{Broadband and Resonant Approaches to
  Axion Dark Matter Detection},''
  \href{http://dx.doi.org/10.1103/PhysRevLett.117.141801}{{\em Phys. Rev.
  Lett.} {\bfseries 117} no.~14, (2016) 141801},
  \href{http://arxiv.org/abs/1602.01086}{{\ttfamily arXiv:1602.01086
  [hep-ph]}}.

\bibitem{Garcon:2017ixh}
A.~Garcon {\em et~al.}, ``{The cosmic axion spin precession experiment
  (CASPEr): a dark-matter search with nuclear magnetic resonance},''
  \href{http://dx.doi.org/10.1088/2058-9565/aa9861}{{\em Quantum Sci. Technol.}
  {\bfseries 3} no.~1, (2017) 014008},
  \href{http://arxiv.org/abs/1707.05312}{{\ttfamily arXiv:1707.05312
  [physics.ins-det]}}.

\bibitem{Crescini:2017uxs}
N.~Crescini, C.~Braggio, G.~Carugno, P.~Falferi, A.~Ortolan, and G.~Ruoso,
  ``{Improved constraints on monopole-dipole interaction mediated by
  pseudo-scalar bosons},''
  \href{http://dx.doi.org/10.1016/j.physletb.2017.09.019}{{\em Phys. Lett. B}
  {\bfseries 773} (2017) 677--680},
  \href{http://arxiv.org/abs/1705.06044}{{\ttfamily arXiv:1705.06044
  [hep-ex]}}.

\bibitem{Lee:2018vaq}
J.~Lee, A.~Almasi, and M.~Romalis, ``{Improved Limits on Spin-Mass
  Interactions},'' \href{http://dx.doi.org/10.1103/PhysRevLett.120.161801}{{\em
  Phys. Rev. Lett.} {\bfseries 120} no.~16, (2018) 161801},
  \href{http://arxiv.org/abs/1801.02757}{{\ttfamily arXiv:1801.02757
  [hep-ex]}}.

\bibitem{Chu:2018dgp}
P.~H. Chu, L.~D. Duffy, Y.~J. Kim, and I.~M. Savukov, ``{Sensitivity of
  Proposed Search for Axion-induced Magnetic Field using Optically Pumped
  Magnetometers},'' \href{http://dx.doi.org/10.1103/PhysRevD.97.072011}{{\em
  Phys. Rev. D} {\bfseries 97} no.~7, (2018) 072011},
  \href{http://arxiv.org/abs/1802.01721}{{\ttfamily arXiv:1802.01721
  [physics.ins-det]}}.

\bibitem{Kim:2021eye}
D.~Kim, Y.~Kim, Y.~K. Semertzidis, Y.~C. Shin, and W.~Yin, ``{Cosmic axion
  force},'' \href{http://dx.doi.org/10.1103/PhysRevD.104.095010}{{\em Phys.
  Rev. D} {\bfseries 104} no.~9, (2021) 095010},
  \href{http://arxiv.org/abs/2105.03422}{{\ttfamily arXiv:2105.03422
  [hep-ph]}}.

\bibitem{Bloch:2022kjm}
{\bfseries NASDUCK} Collaboration, I.~M. Bloch, R.~Shaham, Y.~Hochberg,
  E.~Kuflik, T.~Volansky, and O.~Katz, ``{Constraints on axion-like dark matter
  from a SERF comagnetometer},''
  \href{http://dx.doi.org/10.1038/s41467-023-41162-4}{{\em Nature Commun.}
  {\bfseries 14} no.~1, (2023) 5784},
  \href{http://arxiv.org/abs/2209.13588}{{\ttfamily arXiv:2209.13588
  [hep-ph]}}.

\bibitem{Agrawal:2023lmw}
P.~Agrawal, N.~R. Hutzler, D.~E. Kaplan, S.~Rajendran, and M.~Reig,
  ``{Searching for axion forces with spin precession in atoms and molecules},''
  \href{http://dx.doi.org/10.1007/JHEP07(2024)133}{{\em JHEP} {\bfseries 07}
  (2024) 133}, \href{http://arxiv.org/abs/2309.10023}{{\ttfamily
  arXiv:2309.10023 [hep-ph]}}.

\bibitem{Raffelt:1994ry}
G.~Raffelt and A.~Weiss, ``{Red giant bound on the axion - electron coupling
  revisited},'' \href{http://dx.doi.org/10.1103/PhysRevD.51.1495}{{\em Phys.
  Rev. D} {\bfseries 51} (1995) 1495--1498},
  \href{http://arxiv.org/abs/hep-ph/9410205}{{\ttfamily arXiv:hep-ph/9410205}}.

\bibitem{Capozzi:2020cbu}
F.~Capozzi and G.~Raffelt, ``{Axion and neutrino bounds improved with new
  calibrations of the tip of the red-giant branch using geometric distance
  determinations},'' \href{http://dx.doi.org/10.1103/PhysRevD.102.083007}{{\em
  Phys. Rev. D} {\bfseries 102} no.~8, (2020) 083007},
  \href{http://arxiv.org/abs/2007.03694}{{\ttfamily arXiv:2007.03694
  [astro-ph.SR]}}.

\bibitem{Straniero:2020iyi}
O.~Straniero, C.~Pallanca, E.~Dalessandro, I.~Dominguez, F.~R. Ferraro,
  M.~Giannotti, A.~Mirizzi, and L.~Piersanti, ``{The RGB tip of galactic
  globular clusters and the revision of the axion-electron coupling bound},''
  \href{http://dx.doi.org/10.1051/0004-6361/202038775}{{\em Astron. Astrophys.}
  {\bfseries 644} (2020) A166},
  \href{http://arxiv.org/abs/2010.03833}{{\ttfamily arXiv:2010.03833
  [astro-ph.SR]}}.

\bibitem{Carenza:2021osu}
P.~Carenza and G.~Lucente, ``{Revisiting axion-electron bremsstrahlung emission
  rates in astrophysical environments},''
  \href{http://dx.doi.org/10.1103/PhysRevD.103.123024}{{\em Phys. Rev. D}
  {\bfseries 103} no.~12, (2021) 123024},
  \href{http://arxiv.org/abs/2104.09524}{{\ttfamily arXiv:2104.09524
  [hep-ph]}}.

\bibitem{Adelberger:2009zz}
E.~G. Adelberger, J.~H. Gundlach, B.~R. Heckel, S.~Hoedl, and S.~Schlamminger,
  ``{Torsion balance experiments: A low-energy frontier of particle physics},''
  \href{http://dx.doi.org/10.1016/j.ppnp.2008.08.002}{{\em Prog. Part. Nucl.
  Phys.} {\bfseries 62} (2009) 102--134}.

\bibitem{Moody:1984ba}
J.~E. Moody and F.~Wilczek, ``{NEW MACROSCOPIC FORCES?},''
  \href{http://dx.doi.org/10.1103/PhysRevD.30.130}{{\em Phys. Rev. D}
  {\bfseries 30} (1984) 130}.

\bibitem{Dobrescu:2006au}
B.~A. Dobrescu and I.~Mocioiu, ``{Spin-dependent macroscopic forces from new
  particle exchange},''
  \href{http://dx.doi.org/10.1088/1126-6708/2006/11/005}{{\em JHEP} {\bfseries
  11} (2006) 005}, \href{http://arxiv.org/abs/hep-ph/0605342}{{\ttfamily
  arXiv:hep-ph/0605342}}.

\bibitem{Bluhm:2001rw}
R.~Bluhm, V.~A. Kostelecky, C.~D. Lane, and N.~Russell, ``{Clock comparison
  tests of Lorentz and CPT symmetry in space},''
  \href{http://dx.doi.org/10.1103/PhysRevLett.88.090801}{{\em Phys. Rev. Lett.}
  {\bfseries 88} (2002) 090801},
  \href{http://arxiv.org/abs/hep-ph/0111141}{{\ttfamily arXiv:hep-ph/0111141}}.

\bibitem{Kostelecky:2002hh}
V.~A. Kostelecky and M.~Mewes, ``{Signals for Lorentz violation in
  electrodynamics},'' \href{http://dx.doi.org/10.1103/PhysRevD.66.056005}{{\em
  Phys. Rev. D} {\bfseries 66} (2002) 056005},
  \href{http://arxiv.org/abs/hep-ph/0205211}{{\ttfamily arXiv:hep-ph/0205211}}.

\bibitem{Bluhm:2003un}
R.~Bluhm, V.~A. Kostelecky, C.~D. Lane, and N.~Russell, ``{Probing Lorentz and
  CPT violation with space based experiments},''
  \href{http://dx.doi.org/10.1103/PhysRevD.68.125008}{{\em Phys. Rev. D}
  {\bfseries 68} (2003) 125008},
  \href{http://arxiv.org/abs/hep-ph/0306190}{{\ttfamily arXiv:hep-ph/0306190}}.

\bibitem{Hunter:2013hza}
L.~Hunter, J.~Gordon, S.~Peck, D.~Ang, and J.~F. Lin, ``{Using the Earth as a
  Polarized Electron Source to Search for Long-Range Spin-Spin Interactions},''
  \href{http://dx.doi.org/10.1126/science.1227460}{{\em Science} {\bfseries
  339} no.~6122, (2013) 928--932}.

\bibitem{Clayburn:2024zxx}
N.~B. Clayburn, A.~Glassford, A.~Leiker, T.~Uelmen, J.-F. Lin, and L.~R.
  Hunter, ``{Spherically symmetric Earth models yield no net electron spin},''
  \href{http://dx.doi.org/10.1103/PhysRevD.111.015015}{{\em Phys. Rev. D}
  {\bfseries 111} no.~1, (2025) 015015},
  \href{http://arxiv.org/abs/2411.08050}{{\ttfamily arXiv:2411.08050
  [hep-ph]}}.

\bibitem{Poddar:2023bgk}
T.~K. Poddar and D.~Pachhar, ``{Constraints on monopole-dipole potential from
  tests of gravity},''
  \href{http://dx.doi.org/10.1103/PhysRevD.108.103024}{{\em Phys. Rev. D}
  {\bfseries 108} no.~10, (2023) 103024},
  \href{http://arxiv.org/abs/2302.03882}{{\ttfamily arXiv:2302.03882
  [hep-ph]}}.

\bibitem{Dziewonski:1981xy}
A.~M. Dziewonski and D.~L. Anderson, ``{Preliminary reference earth model},''
  \href{http://dx.doi.org/10.1016/0031-9201(81)90046-7}{{\em Phys. Earth
  Planet. Interiors} {\bfseries 25} (1981) 297--356}.

\bibitem{astronomicalalmanac2024}
{United States Naval Observatory} and {Her Majesty's Nautical Almanac Office},
  {\em The Astronomical Almanac for the Year 2024}.
\newblock U.S. Government Publishing Office and The Stationery Office,
  Washington, DC and London, 2024.

\bibitem{sunwiki}
``{Sun}.'' \url{https://en.wikipedia.org/wiki/Sun}.

\bibitem{earthwiki}
``{Earth}.'' \url{https://en.wikipedia.org/wiki/Earth}.

\bibitem{Dzuba:2024pri}
V.~A. Dzuba, V.~V. Flambaum, and A.~J. Mansour, ``{Constraints on the variation
  of physical constants, equivalence principle violation, and a fifth force
  from atomic experiments},''
  \href{http://dx.doi.org/10.1103/PhysRevD.110.055022}{{\em Phys. Rev. D}
  {\bfseries 110} no.~5, (2024) 055022},
  \href{http://arxiv.org/abs/2402.09643}{{\ttfamily arXiv:2402.09643
  [hep-ph]}}.

\bibitem{Afach:2021pfd}
S.~Afach {\em et~al.}, ``{Search for topological defect dark matter with a
  global network of optical magnetometers},''
  \href{http://dx.doi.org/10.1038/s41567-021-01393-y}{{\em Nature Phys.}
  {\bfseries 17} no.~12, (2021) 1396--1401},
  \href{http://arxiv.org/abs/2102.13379}{{\ttfamily arXiv:2102.13379
  [astro-ph.CO]}}.

\bibitem{Planck:2018vyg}
{\bfseries Planck} Collaboration, N.~Aghanim {\em et~al.}, ``{Planck 2018
  results. VI. Cosmological parameters},''
  \href{http://dx.doi.org/10.1051/0004-6361/201833910}{{\em Astron. Astrophys.}
  {\bfseries 641} (2020) A6}, \href{http://arxiv.org/abs/1807.06209}{{\ttfamily
  arXiv:1807.06209 [astro-ph.CO]}}. [Erratum: Astron.Astrophys. 652, C4
  (2021)].

\bibitem{uniwiki}
``{Universe}.'' \url{https://en.wikipedia.org/wiki/Universe}.

\bibitem{Raffelt:1996wa}
G.~G. Raffelt, {\em Stars as Laboratories for Fundamental Physics: The
  Astrophysics of Neutrinos, Axions, and Other Weakly Interacting Particles}.
\newblock University Of Chicago Press, 1996.
\newblock \url{https://wwwth.mpp.mpg.de/members/raffelt/mypapers/Stars.pdf}.

\bibitem{Raffelt:2006cw}
G.~G. Raffelt, ``{Astrophysical axion bounds},''
  \href{http://dx.doi.org/10.1007/978-3-540-73518-2_3}{{\em Lect. Notes Phys.}
  {\bfseries 741} (2008) 51--71},
  \href{http://arxiv.org/abs/hep-ph/0611350}{{\ttfamily arXiv:hep-ph/0611350}}.

\bibitem{Bottaro:2023gep}
S.~Bottaro, A.~Caputo, G.~Raffelt, and E.~Vitagliano, ``{Stellar limits on
  scalars from electron-nucleus bremsstrahlung},''
  \href{http://dx.doi.org/10.1088/1475-7516/2023/07/071}{{\em JCAP} {\bfseries
  07} (2023) 071}, \href{http://arxiv.org/abs/2303.00778}{{\ttfamily
  arXiv:2303.00778 [hep-ph]}}.

\bibitem{Kippenhahn:2012qhp}
R.~Kippenhahn, A.~Weigert, and A.~Weiss,
  \href{http://dx.doi.org/10.1007/978-3-642-30304-3}{{\em {Stellar structure
  and evolution}}}.
\newblock Astronomy and Astrophysics Library. Springer, 8, 2012.

\bibitem{1970AcA....20...47P}
B.~{Paczy{\'n}ski}, ``{Evolution of Single Stars. I. Stellar Evolution from
  Main Sequence to White Dwarf or Carbon Ignition},'' {\em Acta Astronomica}
  {\bfseries 20} (Jan., 1970) 47.

\bibitem{Caputo:2024oqc}
A.~Caputo and G.~Raffelt, ``{Astrophysical Axion Bounds: The 2024 Edition},''
  \href{http://dx.doi.org/10.22323/1.454.0041}{{\em PoS} {\bfseries
  COSMICWISPers} (2024) 041}, \href{http://arxiv.org/abs/2401.13728}{{\ttfamily
  arXiv:2401.13728 [hep-ph]}}.

\bibitem{Carenza:2024ehj}
P.~Carenza, M.~Giannotti, J.~Isern, A.~Mirizzi, and O.~Straniero, ``{Axion
  Astrophysics},'' \href{http://arxiv.org/abs/2411.02492}{{\ttfamily
  arXiv:2411.02492 [hep-ph]}}.

\bibitem{Catelan:1995ba}
M.~Catelan, J.~A.~d. Freitas~Pacheco, and J.~E. Horvath, ``{The helium-core
  mass at the helium flash in low-mass red giant stars: observations and
  theory},'' \href{http://dx.doi.org/10.1086/177051}{{\em Astrophys. J.}
  {\bfseries 461} (1996) 231},
  \href{http://arxiv.org/abs/astro-ph/9509062}{{\ttfamily
  arXiv:astro-ph/9509062}}.

\bibitem{Raffelt:1985nk}
G.~G. Raffelt, ``{Astrophysical axion bounds diminished by screening
  effects},'' \href{http://dx.doi.org/10.1103/PhysRevD.33.897}{{\em Phys. Rev.
  D} {\bfseries 33} (1986) 897}.

\bibitem{Kublbeck:1990xc}
J.~Kublbeck, M.~Bohm, and A.~Denner, ``{Feyn Arts: Computer Algebraic
  Generation of Feynman Graphs and Amplitudes},''
  \href{http://dx.doi.org/10.1016/0010-4655(90)90001-H}{{\em Comput. Phys.
  Commun.} {\bfseries 60} (1990) 165--180}.

\bibitem{Shtabovenko:2016sxi}
V.~Shtabovenko, R.~Mertig, and F.~Orellana, ``{New Developments in FeynCalc
  9.0}'' \href{http://dx.doi.org/10.1016/j.cpc.2016.06.008}{{\em Comput. Phys.
  Commun.} {\bfseries 207} (2016) 432--444},
  \href{http://arxiv.org/abs/1601.01167}{{\ttfamily arXiv:1601.01167
  [hep-ph]}}.

\bibitem{Shtabovenko:2020gxv}
V.~Shtabovenko, R.~Mertig, and F.~Orellana, ``{FeynCalc 9.3: New features and
  improvements},'' \href{http://dx.doi.org/10.1016/j.cpc.2020.107478}{{\em
  Comput. Phys. Commun.} {\bfseries 256} (2020) 107478},
  \href{http://arxiv.org/abs/2001.04407}{{\ttfamily arXiv:2001.04407
  [hep-ph]}}.

\bibitem{github}
P.~Carenza. \url{https://github.com/pcarenza95/LIVBremsstrahlung}.

\bibitem{Raffelt:1989zt}
G.~G. Raffelt, ``{Axion bremsstrahlung in red giants},''
  \href{http://dx.doi.org/10.1103/PhysRevD.41.1324}{{\em Phys. Rev. D}
  {\bfseries 41} (1990) 1324--1326}.

\bibitem{Hardy:2016kme}
E.~Hardy and R.~Lasenby, ``{Stellar cooling bounds on new light particles:
  plasma mixing effects},''
  \href{http://dx.doi.org/10.1007/JHEP02(2017)033}{{\em JHEP} {\bfseries 02}
  (2017) 033}, \href{http://arxiv.org/abs/1611.05852}{{\ttfamily
  arXiv:1611.05852 [hep-ph]}}.

\bibitem{1939isss.book.....C}
S.~{Chandrasekhar}, {\em {An introduction to the study of stellar structure}}.
\newblock 1939.

\bibitem{Arvanitaki:2009fg}
A.~Arvanitaki, S.~Dimopoulos, S.~Dubovsky, N.~Kaloper, and J.~March-Russell,
  ``{String Axiverse},''
  \href{http://dx.doi.org/10.1103/PhysRevD.81.123530}{{\em Phys. Rev. D}
  {\bfseries 81} (2010) 123530},
  \href{http://arxiv.org/abs/0905.4720}{{\ttfamily arXiv:0905.4720 [hep-th]}}.

\bibitem{Arvanitaki:2010sy}
A.~Arvanitaki and S.~Dubovsky, ``{Exploring the String Axiverse with Precision
  Black Hole Physics},''
  \href{http://dx.doi.org/10.1103/PhysRevD.83.044026}{{\em Phys. Rev. D}
  {\bfseries 83} (2011) 044026},
  \href{http://arxiv.org/abs/1004.3558}{{\ttfamily arXiv:1004.3558 [hep-th]}}.

\bibitem{Arvanitaki:2014wva}
A.~Arvanitaki, M.~Baryakhtar, and X.~Huang, ``{Discovering the QCD Axion with
  Black Holes and Gravitational Waves},''
  \href{http://dx.doi.org/10.1103/PhysRevD.91.084011}{{\em Phys. Rev. D}
  {\bfseries 91} no.~8, (2015) 084011},
  \href{http://arxiv.org/abs/1411.2263}{{\ttfamily arXiv:1411.2263 [hep-ph]}}.

\bibitem{Stott:2018opm}
M.~J. Stott and D.~J.~E. Marsh, ``{Black hole spin constraints on the mass
  spectrum and number of axionlike fields},''
  \href{http://dx.doi.org/10.1103/PhysRevD.98.083006}{{\em Phys. Rev. D}
  {\bfseries 98} no.~8, (2018) 083006},
  \href{http://arxiv.org/abs/1805.02016}{{\ttfamily arXiv:1805.02016
  [hep-ph]}}.

\bibitem{Baryakhtar:2020gao}
M.~Baryakhtar, M.~Galanis, R.~Lasenby, and O.~Simon, ``{Black hole
  superradiance of self-interacting scalar fields},''
  \href{http://dx.doi.org/10.1103/PhysRevD.103.095019}{{\em Phys. Rev. D}
  {\bfseries 103} no.~9, (2021) 095019},
  \href{http://arxiv.org/abs/2011.11646}{{\ttfamily arXiv:2011.11646
  [hep-ph]}}.

\bibitem{Mehta:2021pwf}
V.~M. Mehta, M.~Demirtas, C.~Long, D.~J.~E. Marsh, L.~McAllister, and M.~J.
  Stott, ``{Superradiance in string theory},''
  \href{http://dx.doi.org/10.1088/1475-7516/2021/07/033}{{\em JCAP} {\bfseries
  07} (2021) 033}, \href{http://arxiv.org/abs/2103.06812}{{\ttfamily
  arXiv:2103.06812 [hep-th]}}.

\bibitem{Hoof:2024quk}
S.~Hoof, D.~J.~E. Marsh, J.~Sisk-Reyn{\'e}s, J.~H. Matthews, and C.~Reynolds,
  ``{Getting More Out of Black Hole Superradiance: a Statistically Rigorous
  Approach to Ultralight Boson Constraints},''
  \href{http://arxiv.org/abs/2406.10337}{{\ttfamily arXiv:2406.10337
  [hep-ph]}}.

\bibitem{Lambiase:2025twn}
G.~Lambiase, T.~K. Poddar, and L.~Visinelli, ``{Impact of the cosmic neutrino
  background on black hole superradiance},''
  \href{http://arxiv.org/abs/2503.02940}{{\ttfamily arXiv:2503.02940
  [hep-ph]}}.

\bibitem{Brito:2014wla}
R.~Brito, V.~Cardoso, and P.~Pani, ``{Black holes as particle detectors:
  evolution of superradiant instabilities},''
  \href{http://dx.doi.org/10.1088/0264-9381/32/13/134001}{{\em Class. Quant.
  Grav.} {\bfseries 32} no.~13, (2015) 134001},
  \href{http://arxiv.org/abs/1411.0686}{{\ttfamily arXiv:1411.0686 [gr-qc]}}.

\bibitem{Hui:2016ltb}
L.~Hui, J.~P. Ostriker, S.~Tremaine, and E.~Witten, ``{Ultralight scalars as
  cosmological dark matter},''
  \href{http://dx.doi.org/10.1103/PhysRevD.95.043541}{{\em Phys. Rev. D}
  {\bfseries 95} no.~4, (2017) 043541},
  \href{http://arxiv.org/abs/1610.08297}{{\ttfamily arXiv:1610.08297
  [astro-ph.CO]}}.

\bibitem{Hubisz:2024hyz}
J.~Hubisz, S.~Ironi, G.~Perez, and R.~Rosenfeld, ``{A note on the quality of
  dilatonic ultralight dark matter},''
  \href{http://dx.doi.org/10.1016/j.physletb.2024.138583}{{\em Phys. Lett. B}
  {\bfseries 851} (2024) 138583},
  \href{http://arxiv.org/abs/2401.08737}{{\ttfamily arXiv:2401.08737
  [hep-ph]}}.

\bibitem{Banerjee:2022wzk}
A.~Banerjee, J.~Eby, and G.~Perez, ``{From axion quality and naturalness
  problems to a high-quality ZN QCD relaxion},''
  \href{http://dx.doi.org/10.1103/PhysRevD.107.115011}{{\em Phys. Rev. D}
  {\bfseries 107} no.~11, (2023) 115011},
  \href{http://arxiv.org/abs/2210.05690}{{\ttfamily arXiv:2210.05690
  [hep-ph]}}.

\bibitem{Giudice:2016yja}
G.~F. Giudice and M.~McCullough, ``{A Clockwork Theory},''
  \href{http://dx.doi.org/10.1007/JHEP02(2017)036}{{\em JHEP} {\bfseries 02}
  (2017) 036}, \href{http://arxiv.org/abs/1610.07962}{{\ttfamily
  arXiv:1610.07962 [hep-ph]}}.

\bibitem{Wood:2023lis}
K.~Wood, P.~M. Saffin, and A.~Avgoustidis, ``{Clockwork cosmology},''
  \href{http://dx.doi.org/10.1088/1475-7516/2023/07/062}{{\em JCAP} {\bfseries
  07} (2023) 062}, \href{http://arxiv.org/abs/2304.09205}{{\ttfamily
  arXiv:2304.09205 [hep-th]}}.

\bibitem{Batell:2022qvr}
B.~Batell, A.~Ghalsasi, and M.~Rai, ``{Dynamics of dark matter misalignment
  through the Higgs portal},''
  \href{http://dx.doi.org/10.1007/JHEP01(2024)038}{{\em JHEP} {\bfseries 01}
  (2024) 038}, \href{http://arxiv.org/abs/2211.09132}{{\ttfamily
  arXiv:2211.09132 [hep-ph]}}.

\bibitem{Cong:2024qly}
L.~Cong {\em et~al.}, ``{Spin-dependent exotic interactions},''
  \href{http://arxiv.org/abs/2408.15691}{{\ttfamily arXiv:2408.15691
  [hep-ph]}}.

\bibitem{OHare:2020wah}
C.~A.~J. O'Hare and E.~Vitagliano, ``{Cornering the axion with $CP$-violating
  interactions},'' \href{http://dx.doi.org/10.1103/PhysRevD.102.115026}{{\em
  Phys. Rev. D} {\bfseries 102} no.~11, (2020) 115026},
  \href{http://arxiv.org/abs/2010.03889}{{\ttfamily arXiv:2010.03889
  [hep-ph]}}.

\end{thebibliography}\endgroup

\end{document}